\documentclass[
        parskip = half, 
        twoside,
        footinclude = false, 
        ]{scrartcl}
\usepackage[T1]{fontenc}     
\usepackage[utf8]{inputenc}
\usepackage[left=30mm,right=25mm,textheight=245mm,top=25mm,footskip=10mm]{geometry} 
\usepackage[ngerman,english]{babel} 
\usepackage[autostyle]{csquotes}    
\AtBeginDocument{\languageshorthands{ngerman}\useshorthands{"}} 
\usepackage{fancyhdr}  
\usepackage{graphicx}  
\usepackage{color}     
\usepackage{psfrag}    
\usepackage{amsmath}   
\usepackage{amssymb}   
\usepackage{ifthen}    
\usepackage{multirow}  
\usepackage{booktabs}  
\usepackage[numbers,sort]{natbib} 

\newboolean{boldeqnitalic}
\setboolean{boldeqnitalic}{true}

\usepackage{defdb1}
\usepackage{defet1}
\usepackage{tikzdef}

\addtokomafont{caption}{\small} 
\setcapindent{0em} 
\setcapwidth{0.85\textwidth}
\setkomafont{captionlabel}{\bfseries} 
\deffootnote[1em]{1.0em}{1em}{\textsuperscript{\thefootnotemark}}
\makeatletter
\renewcommand\section{\@startsection
   {section}{1}{0mm}
   {1.0\baselineskip}
   {0.2\baselineskip}
   {\normalfont\bfseries\Large}
   }
\makeatother
\makeatletter
\renewcommand\subsection{\@startsection
   {subsection}{2}{0mm}
   {0.5\baselineskip}
   {0.1\baselineskip}
   {\normalfont\bfseries\large}
   }
\makeatother
\makeatletter
\renewcommand\subsubsection{\@startsection
   {subsubsection}{3}{0mm}
   {0.5\baselineskip}
   {-1em}
   {\normalfont\bfseries\large}
   }
\makeatother
\definecolor{tudresden_blue}{rgb}{0.043,0.164,0.316}

\definecolor{graph_blue}{RGB}{57,106,177} 
\definecolor{graph_orange}{RGB}{218,124,48}
\definecolor{graph_green}{RGB}{62,150,81}
\definecolor{graph_red}{RGB}{204,37,41}
\definecolor{graph_gray}{RGB}{83,81,84}
\definecolor{graph_purple}{RGB}{107,76,154}
\definecolor{graph_darkred}{RGB}{146,36,40}
\definecolor{graph_greenish}{RGB}{148,139,61}

\definecolor{rub_green}{cmyk}{0.75,0.0,1.0,0}      

\colorlet{colorA}{orange}
\colorlet{colorB}{blue}

\pgfplotscreateplotcyclelist{style-colors}{%
	graph_blue,every mark/.append style={solid,fill=.!80!black},mark=square*\\%
	graph_orange,every mark/.append style={solid,fill=.!80!black},mark=*\\%
	rub_green,every mark/.append style={solid,fill=.!80!black},mark=otimes*\\%
	graph_red,mark=star\\%
	graph_gray,every mark/.append style={solid,fill=.!80!black},mark=diamond*\\%
	graph_purple,every mark/.append style={solid,fill=.!80!black},mark=*\\%
	graph_darkred,\\%
	graph_greenish\\%
	every mark/.append style={solid,fill=.!80!black},mark=square*\\%
}

\newcommand{\figpath}{figures}       
\newcommand{\tikzpath}{diagrams}

\usetikzlibrary{arrows.meta} 

\usepackage[linesnumberedhidden,vlined,plain,noend,figure]{algorithm2e}

\SetAlgoNoEnd
\SetAlgoNoLine%
\SetKwComment{eq}{Eq. }{}
\definecolor{comment_darkgray}{rgb}{0.2, 0.2, 0.2}

\SetCommentSty{mycomment}
\SetKwFor{Loop}{loop}{}{}%
\SetKwInput{Input}{input}
\SetKwInput{bc}{boundary conditions}
\SetKw{Compute}{compute}
\SetKw{Solve}{solve}
\SetKw{Update}{update}

\usepackage{caption}

\usepackage{subcaption}

\begin{document}


\clearpage
\setcounter{page}{1}

\begin{center}
{\LARGE A General, Implicit, Large-Strain FE$^2$ Framework for the Simulation of Dynamic Problems on Two Scales}

\vspace{5mm}

Erik Tamsen$^{1,2\star}$, Daniel Balzani$^{2}$

\vspace{3mm}

{\small $^1$ Institute of Mechanics and Shell Structures, Technische Universität Dresden, 01062 Dresden}\\

{\small $^2$Chair of Continuum Mechanics, Ruhr University Bochum, 44801 Bochum}\\

\vspace{3mm}

{\small ${}^{\star}$Email address of corresponding author: erik.tamsen@tu-dresden.de}

\vspace{10mm}

\begin{minipage}{15.0cm}
\textbf{Abstract}\hspace{3mm}

In this paper we present a fully-coupled, two-scale homogenization method for dynamic loading in the spirit of FE$^2$ methods. 
The framework considers the balance of linear momentum including inertia at the microscale to capture possible dynamic effects arising from micro heterogeneities. 
A finite-strain formulation is adapted to account for geometrical nonlinearities enabling the study of e.g. plasticity or fiber pullout, which may be associated with large deformations. 
A consistent kinematic scale link is established as displacement constraint on the whole representative volume element. 
The consistent macroscopic material tangent moduli are derived including micro inertia in closed form. 
These can easily be calculated {with} a loop over all microscopic finite elements, only applying existing assembly and solving procedures. Thus, making it suitable for standard {finite element} program architectures. 
Numerical examples of a layered periodic material are presented and compared to direct numerical simulations to demonstrate the capability of the proposed framework. 

\end{minipage}
\end{center}

\medskip{}


\section{Introduction}
Micro-heterogeneous materials can give rise to wave propagation under dynamic loading, leading to distinct stress distributions at the microscale. Thus, resulting in a complex macroscopic material behavior. 
There are various examples from different fields of application which cover a broad range of length scales. 
Currently the interest is high in metamaterials in general, but especially in locally resonant metamaterials exhibiting special properties like band gaps and negative bulk moduli. 
The applications range from cloaking devices \citep{kadi:2015:eoci,Cumm:2007:Opa} over tunable sound attenuation \citep{Liu:2000:s, Li:2004:pre} to earthquake protection \citep{brul:2014:eosm, mini:2016:lsmm}. 
More classical materials are being investigated as well. 
One example is metaconcrete, which replaces aggregates by rubber-coated lead inclusions to weaken impact waves \citep{mitc:2016:eoff, Kett:2018:Eid}. 
A different approach for an improved impact resistance {are} strain-hardening cement-based composites (SHCC), that show a pronounced energy dissipation under dynamic loading, as well as a change in fiber failure and overall crack pattern \citep{Curo:2016:eofp,curo:2017:povs,curo:2019:mbcf}. 
In addition to that, porous materials have shown an influence of microscopic inertia on voids under high strain rates \citep{Moli:2001:mmop, Sart:2015:mom}. 
This list illustrates the possible influence of the material microstructure on the macroscopic response under high dynamic loading for a wide range of materials and applications. 
In general, any material under impact loading, which has large variation in stiffness, e.g. rubber-coated particles, or materials with a pronounced variation in density, e.g. if pores or cracks are present at the microscale, can exhibit {distinct} effective macroscopic properties resulting from the dynamics in the microstructure.\par

To model the before-mentioned effects, the dynamics at the microscale need to be accounted for. 
Computational homogenization methods for quasi-static loading have become a common tool in numerical material analysis, see e.g. {\citep{SmiBreMei:1998:pot, MouSuq:1998:anm, Mieh:1999:cmmt, FisShe:1999:fdp, Fey:1999:mfe, TerHorKyoKik:2000:sot}.} A {recent} overview of computational homogenization methods {in general} is given in~\citep{geer:2017:hmam}. {In addition a recapitulation of the FE$^2$ method in particular is presented in ~\citep{Sch:2013:ant}.}
However, with the rise in interest in metamaterials more and more dynamic homogenization frameworks have been published in the last years. 
One is the method of asymptotic expansion, see e.g., \citep{Fish:2002:nldm,cras:2010:hfhf,hui:2014:ahoh, Hu:2019:Mne}, which is mainly based on the original work of Bensoussan et.~al.~\citep{Bens:1978:aafp}. 
Then there is the more general theory of elastodynamic homogenization by Willis~\citep{Will:1997:doc}, which has been applied in e.g. \mbox{\citep{Milt:2010:potr,Nema:2011:jotm,Will:2012:crm}} and others. 
The two approaches are based on related ideas and their similarities are studied in~\citep{nass:2016:oaeh}. 
Both methods are limited to elastic, periodic media. 
A more general approach is the micro-macro simulation based on a representative volume element (RVE). 
In such homogenization methods, macroscopic quantities such as the deformation gradient and the displacement vector at a macroscopic integration point are projected onto a microscopic boundary value problem whose homogenized response replaces the constitutive material law. 
In the case where the finite element method (FEM) is used on both scales, this procedure is called the FE$^2$ method. 
A good theoretic introduction to this theory including dynamics is given by \citep{Souz:2015:arbm}, which has been the foundation on which this paper is built. 
The framework in \citep{Fish:2002:nldm} calculates a quasi-static microstructure but then applies an inertia-induced eigenstrain based on the microstructure as an extra body force at the macroscale to account for micro-inertia effects. This was extended by \citep{kara:2014:adms} to account for matrix cracking at the microscale under impact loading. 
Other, rather FE$^2$-type schemes as \citep{liu:2017:vcgp,liu:2018:dhor,Pham:2013:tchf,Roca:2018:acmh,srid:2018:agmf,wang:2002:mmii} calculate the full balance of linear momentum at the microscale. 
In \citep{liu:2017:vcgp} an explicit, periodic, small-strain framework is presented, which was extended to an implicit time integration method for modeling resonant elastic metamaterials in \citep{liu:2018:dhor}.  \citep{Pham:2013:tchf,Roca:2018:acmh} use the assumption of linear elasticity to improve the computational performance, by splitting the problem into a purely static and a special dynamic boundary value problem (BVP). 
To better capture a wider range of applied frequencies, \citep{srid:2018:agmf} use a Floquet-Bloch transformation to build a base of eigenmodes to analyze elastic, periodic metamaterials. 
The mentioned frameworks all use at least one of the following idealizations: small strains, linear elasticity, periodic or symmetric microstructures. 
In addition, many require quite elaborate implementations.\par

The aim of this paper is to build a multiscale framework for dynamic loading as general as possible, while still being compatible with standard FE architecture. 
To enable the analysis of micro-mechanical processes as plasticity or fiber pullout, as well as to incorporate effects resulting from geometric nonlinearities, the proposed framework uses a finite-strain formulation. 
Its importance is supported by the analytical example in \citep{khaj:2014:dcoa} of a nonlinear, elastic metamaterial, where finite strains were shown to be relevant for large wave amplitudes. 
To permit a flexible damage evolution in the RVE, which is not dominated by periodic boundary conditions \citep{Coen:2012:nbcf}, we propose to apply kinematic scale links as constraints on the whole RVE. 
This allows us to model any type of RVE morphology. 
\par

The paper starts by discussing the general ideas of the framework in \mbox{Section \ref{sec:framework}}. 
The used notations of the large-strain framework are presented and the averaging relations derived in \citep{souz:2010:vfol} are briefly recapitulated to include the full balance of momentum at the microscale. 
Then in \mbox{Section \ref{sec:microscale}} the FE formulations of the microscale are presented and the kinematic constraints are derived. 
In \mbox{Section \ref{sec:macroscale}} the respective macroscale formulations are displayed. 
There, a focus is laid on the derivation of consistent macroscale tangent moduli in closed form. These enable a quadratically converging macroscopic Newton-iteration, resulting in a robust and efficient algorithm compared to numerical differentiation through perturbation of the macroscopic quantities \citep{Mieh:1996:ncoa}. 
Once the whole theoretical background of the framework is explained, \mbox{Section \ref{sec:numericalexamples}} provides numerical examples, demonstrating the convergence behavior and analyzing some properties of RVEs under dynamic loading.  
The paper is completed in \mbox{Section \ref{sec:conclusion}} with concluding remarks.


\section{Homogenization Framework Including Microscopic Inertia} \label{sec:framework}
The general idea of the homogenization framework for dynamics is to consider the full balance of linear momentum including the inertia terms at the microscale. 
The enables not only the analysis of full dynamic fields at the microscale but a direct study of microscopic inertia effects on the macroscale.
By using appropriate averaging relations and kinematic links, a consistent scale bridging for dynamic loading is established. 
In the FE$^2$ method, each macroscopic Gauss point is associated with a separate microscopic RVE simulation which uses the macroscopic mechanical quantities to define the microscopic BVP. 
In order to differentiate the two scales, variables associated with the macroscale are denoted with a bar $\ol{\bullet}$. 
Here, the finite-strain framework is taken into account in order to enable the analysis of a wide range of material behavior and micro-mechanical effects under dynamic loading. 
In the following sections the fundamental ingredients of the scale-coupling framework are explained, see also the schematic illustration in Figure \ref{fig:framework}. 

\begin{figure}[b]
	\begin{minipage}[b]{.5\textwidth}
		\captionsetup{width=0.95\textwidth}
		\centering			
   	\input{\tikzpath/multiscale/micromacro}
   	\caption{Overview of the FE$^2$ framework including microscale dynamics. Here, an example of macroscopic impact on SHCC is illustrated.}
		\label{fig:framework}
	\end{minipage}%
	\hfill
	\begin{minipage}[b]{.45\textwidth}
		\captionsetup{width=\textwidth}
			\begin{subfigure}{0.45\textwidth}
				\centering
				\tikzset{external/export next=false}
\begin{tikzpicture}

\coordinate(A) at (1,1); 

\coordinate(B) at (5,1); 
\coordinate(C) at (3,1); 
\coordinate(D) at (3,1.7); 

\node[circle](AA) at (A){};
\node[circle](BB) at (B){};

\draw [fill=gray!20] (A) ellipse (0.7 and 0.7);
\fill [black] (A) ellipse (0.1 and 0.1);


\draw [fill=gray!20] (B) ellipse (0.9 and 0.5);
\fill [black] (B) ellipse (0.1 and 0.1);

\node[circle,above left] at (A){\footnotesize $\ol{\bX}$};
\node[circle,above right] at (B){\footnotesize $\ol{\bx}$};

\node[above] at (C){\footnotesize $\ol{\bF} = \bone + \ol{\bH}$};
\node[above] at (D){\footnotesize $\ol{\bx} = \ol{\bX} + \ol{\bu}$};

\draw[out=35, in=180-35,-{Latex[length=3mm,angle'=30]},line width=0.2mm] (AA) to (BB);
\draw[out=0, in=180,-{Latex[length=3mm,angle'=30]},line width=0.2mm] (AA) to (BB);

\node (a) at (0.2,1.7){\footnotesize $\ol{\B}$};
\node (a) at (5.7,1.7){\footnotesize $\ol{\Bt}$};

\end{tikzpicture}
				\caption{Macroscale}
			\end{subfigure}\\
			\begin{subfigure}{0.45\textwidth}
				\centering
				\tikzset{external/export next=false}
\begin{tikzpicture}

\coordinate(A) at (1,1); 

\coordinate(B) at (5,1); 
\coordinate(C) at (3,1); 
\coordinate(D) at (3,1.7); 

\node[circle](AA) at (A){};
\node[circle](BB) at (B){};

\draw [fill=gray!20] (A) ellipse (0.7 and 0.7);
\fill [black] (A) ellipse (0.1 and 0.1);


\draw [fill=gray!20] (B) ellipse (0.9 and 0.5);
\fill [black] (B) ellipse (0.1 and 0.1);

\node[circle,above left] at (A){\footnotesize ${\bX}$};
\node[circle,above right] at (B){\footnotesize ${\bx}$};

\node[above] at (C){\footnotesize ${\bF} = \ol{\bF} + \widetilde{\bH}$};
\node[above] at (D){\footnotesize ${\bx} = \ol{\bu} + \ol{\bF}{\bX} + \widetilde{\bu}$};

\draw[out=35, in=180-35,-{Latex[length=3mm,angle'=30]},line width=0.2mm] (AA) to (BB);
\draw[out=0, in=180,-{Latex[length=3mm,angle'=30]},line width=0.2mm] (AA) to (BB);

\node (a) at (0.2,1.7){\footnotesize ${\B}$};
\node (a) at (5.7,1.7){\footnotesize ${\Bt}$};

\end{tikzpicture}
				\caption{Microscale}
			\end{subfigure}
		\caption{Large-strain continuum mechanics on both scales.}
		\label{fig:micromacrocontinuum}
	\end{minipage}
\end{figure}

\subsection{Large-Strain Kinematics at Different Scales} \label{ssec:framework_kinematics}
The connection of the (undeformed) reference and the (deformed) current configuration is described by the displacement $\bu$ as $\bu = \bx - \bX$,  where $\bX\in{\cal B}$ denotes the coordinates in the reference configuration and $\bx\in{\cal S}$ the deformation. 
The link between the two configurations in terms of transformations of vector elements is described by the deformation and displacement gradients, respectively $\bF = \partial_{\bX}{\bx}=\bone+\bH$ and $\bH = \partial_{\bX}{\bu}$, such that $\bx = \bF\bX$. 
In this work, the origin of the microscopic coordinates is chosen as the geometrical center of the RVE, with 
\begin{align}
\int_{\B} \bX \dV = \bzero  ,\label{eq:origin}
\end{align}
where $\int_{\B} \dV$ is the volume integral over the microscopic reference body $\B$. 
Note that this choice of origin has no influence on the results but {it} simplifies the notation. 
Now the microscopic deformation $\bx$ can be split into the sum of terms, 
\begin{align}
\bx = \ol{\bu} + \ol{\bF}\bX + \widetilde{\bu}. \label{eq:deformationsplit}
\end{align} 
Herein, two terms result directly from the macroscale: a constant part $\ol{\bu}$, which describes the macroscopic rigid body translations, and a homogeneous part $\ol{\bF}\bX$, defined in terms of the macroscopic deformation gradient.

The difference of these homogeneous deformations $\ol{\bu}+\ol{\bF}\bX$ to the actual deformations $\bx$ is the microscopic displacement fluctuation field $\widetilde{\bu}$. 
This is the field the microscopic BVP will be solved for. 
Now the microscopic displacements $\bu$ can be written as
\begin{align}
\bu = \ol{\bu} + \ol{\bH}\bX + \widetilde{\bu} . \label{eq:displacementsplit}
\end{align} 
Analogously, the microscopic deformation gradient can be split as 
\begin{align}
\bF = \ol{\bF} + \widetilde{\bH} \quad\text{with}\quad \widetilde{\bH} = \partial_{\bX}\widetilde{\bu}.\label{eq:Fsplit}
\end{align}
At the macroscale the kinematics are standard, see Figure~\ref{fig:micromacrocontinuum}, where the kinematic relations for both the micro- and macroscale are illustrated. 

\subsection{Averaging Relations}
To expand quasi-static homogenization frameworks to the case of dynamics, an extended version of the Hill-Mandel condition of macro homogeneity \citep{Hill:1972:ocmv,Mand:1971:pcev}, which takes into account inertia and body forces at the microscale, is adopted
\begin{align}
  \ol{\bP}:\delta\ol{\bF} - \ol{\bbf} \cdot \delta \ol{\bu} = \left\langle  {\bP}:\delta{\bF} - {\bbf} \cdot \delta {\bu} \right\rangle  ,\label{eq:hillmandel}
\end{align}
see~\citep{Blan:2016:vfag} and~\citep{Souz:2015:arbm}. 
Herein, $\left\langle \bullet \right\rangle  =\frac{1}{V} \int_{\B} \bullet \dV$ is an expression used to abbreviate the volume average of a microscopic quantity. 
$\bP$ is the first Piola-Kirchhoff stress tensor, $\bbf$ is the microscopic body force vector, and $\ol{\bbf}$ is its macroscopic counterpart. 
The variational equation \eqref{eq:hillmandel} is also called the Principle of Multiscale Virtual Power. 
It ensures that the virtual work of the macroscale coincides with its respective microscopic volume average, thus ensuring energetic consistency across the scales. 
By decomposing $\delta\bF$ and $\delta\bu$, following \eqref{eq:displacementsplit} and \eqref{eq:Fsplit}, three important equations can be derived. 
First, one obtains
\begin{align}
\left\langle  {\bP}:\delta\widetilde{\bF} - {\bbf} \cdot \delta\widetilde{\bu}  \right\rangle = 0,
\end{align}
which is automatically fulfilled for mechanical equilibrium. 
Then, more importantly, the averaging equation for the effective macroscopic stress $\ol{\bP}$ and the effective macroscopic body force vector $\ol{\bbf}$ are derived as
\begin{align}
\ol{\bP} &= \left\langle  {\bP}  - {\bbf} \otimes {\bX}  \right\rangle  \quad\text{and}\label{eq:macroP}\\
\ol{\bbf} &= \left\langle {\bbf} \right\rangle\label{eq:macrof}.
\end{align}
These averaging relations can be found in multiple frameworks dealing with homogenization of dynamics, see e.g., \citep{liu:2017:vcgp, Souz:2015:arbm,Pham:2013:tchf,Liu:2016:darf,Roca:2018:acmh,Moli:2001:mmop}. 
It was shown in \citep{Liu:2016:Dar} that the extended Hill-Mandel averaging relation can be applied in a discretized setting without introducing additional error by scale transition. 

\subsection{Scale Separation}
A principal ingredient of the Hill-Mandel condition, as well as the presented extension \eqref{eq:hillmandel}, is the assumption of scale separation. 
This means that the equation only holds if the length scale of the microscopic mechanical fields is significantly smaller than the one of the macroscopic mechanical fields. 
For dynamic homogenization this means in practice, that the macroscopic wavelength is sufficiently larger than the size of the RVE. 

\subsection{Time Integration}
In this paper, an implicit numerical time integration method of first order is applied. 
Considering 
\begin{align}
\ddot{\bullet} = \frac{(\alpha_1{\bullet} - \alpha_2\hat{\bullet})}{\Delta t^{2}} ,
\end{align}
we can express the second time derivative of any quantity $\bullet$ as the difference of the current time step and the last $\hat{\bullet}$, divided by the square of the time step $\Delta t$. 
The parameters $\alpha_1$ and $\alpha_2$ define the specific time integration scheme. 
For an explicit time integration, $\alpha_1$ can be set to zero. 
Applying this to derivatives of acceleration terms with respect to a quantity $\Box$ in the current configuration leads to
\begin{align}
\frac{\partial \ddot{\bullet}}{\partial \Box} =
\frac{1}{\Delta t^{2}}\frac{\partial (\alpha_1{\bullet} - \alpha_2\hat{\bullet})}{\partial \Box } = 
\frac{\alpha_1}{\Delta t^{2}}\frac{\partial {\bullet}}{\partial \Box} .\label{eq:ti1}
\end{align}
Analogously, the derivative of a value $\bullet$ with respect to the second time derivative $\ddot{\Box}$ reads 
\begin{align}
\frac{\partial {\bullet}}{\partial \ddot{\Box}} = \left( 
\frac{\partial \ddot{\Box}}{\partial {\bullet}} \right)^{-1} = 
\left( \frac{1}{\Delta t^{2}}\frac{\alpha_1\partial ({\Box} - \alpha_1 \hat{\Box})}{\partial \bullet } \right)^{-1}= 
\frac{\Delta t^{2}}{\alpha_1}\frac{\partial {\bullet}}{\partial \Box} .\label{eq:ti2}
\end{align}
For the numerical simulations in the numerical analysis section the widely used Newmark method \citep{Newm:1959:joem} is applied, with $\alpha_1 = \frac{1}{\beta}$. 
Herein, $\beta$ is one of the two parameters of the Newmark method influencing the type and stability of the time integration.


\FloatBarrier

\section{The Microscopic Problem} \label{sec:microscale}
We start with the formulation of the microscopic problem, which includes the necessary kinematic links to the macroscale. 
Furthermore, an algorithmic treatment for the associated constraint conditions is given. 

\subsection{Microscopic Balance of Linear Momentum}
For a dynamic analysis, the microscopic balance of linear momentum is given by
\begin{align}
\Div{\bP} + {\bbf} = \bzero.
\end{align}
The body force vector $\bbf$ can be decomposed into an inertia part $\bbf^{\rho}$ and a body force vector representing e.g. the gravitational pull $\bbf^{\text{b}}$.
As this framework is intended to model impact loading, gravitational forces are assumed to be negligible compared to the inertia forces. 
Thus, the {relevant} body force vector is defined as ${\bbf} \coloneqq {\bbf}^{\rho}= -\rho_0 \ddot{\bu}$ with $\rho_{0}$ referring to the density of the microscale constituents in the reference configuration. 
If gravitational forces have to be considered, they can be included in the standard way by the additional force vector $\bbf^{\text{b}} = \rho_0\,\bg$, where $\bg$ is the gravitation field. 
Since this force however, does not depend on the displacements, it does not represent any specialty with view to the proposed homogenization framework and is thus omitted from the presentation to avoid unnecessary complications. \\
Using standard FE procedures for the discretization and linearization of the weak form of the balance of linear momentum, the well-known equation 
\begin{align}
\delta \widetilde{\bD}^{\text{T}}\widehat{\bK} \Delta \widetilde{\bD} = \delta \widetilde{\bD}^{\text{T}}\widehat{\bR} \label{eq:soe}
\end{align}
is obtained, where $\widehat{\bK}$ and $\widehat{\bR}$ are the global tangent stiffness matrix and the residuum matrix, respectively. 
In the following, the hat~$\widehat{\bullet}$ is used to highlight quantities that include dynamic terms.
After incorporating the Dirichlet boundary conditions, the global matrix of incremental nodal displacements~$\Delta\widetilde{\bD}$ are computed from $\widehat{\bK}\Delta\widetilde{\bD}=\widehat{\bR}$ in each Newton iteration step in order to obtain the updated displacements until convergence of the Newton scheme is achieved, i.e. until $|\Delta\widetilde{\bD}|<tol$. 
For the classical scheme, the global tangent stiffness matrix $\widehat{\bK}$ is assembled from the element matrix defined as 
\begin{align}
	&\widehat{\bk}^e  = {\bk}^e + \frac{{\alpha}_1}{ \Delta t^2}{\bbm}^e \text{,\quad with}  \label{eq:mickele}\\
{\bk}^e = \int_{\B^e} {\bB}^{e^{\text{T}}}\! &{\IA} {\bB}^e   \dV    \quad \text{and} \quad
{\bbm}^e = \int_{\B^e} {\bN}^e\! \rho_0 {\bN}^{e^{\text{T}}} \dV .
\end{align}
Herein, $\bN^e$ is the classical element matrix of shape functions, $\bB^e$ denotes the classical B-matrix containing the derivatives of the shape functions, and $\IA$ is the matrix representation of the material tangent modulus, defined as the sensitivity of the microscopic stress with respect to the microscopic deformation gradient as $\IA = \partial_{{\bF}}{\bP}$. 
Analogously, the global residuum matrix {$\widehat{\bR}$} is obtained {by the assembly of} the element-wise counterparts, given as 
\begin{align}
\widehat{\bbr}^e = \int_{\B^e} \left(   {\bB}^{e^{\text{T}}}\! \bP
+ {\bN}^e\! \rho_0 \ddot{\bu} \right)  \dV   . \label{eq:micrele}
\end{align}
Herein, $\bP$ denotes the matrix representation of the {first} Piola-Kirchhoff stresses. 
It can be noted that both, $\widehat{\bk}^e$ and $\widehat{\bbr}^e$ have dynamic terms related to the density $\rho_0$, which directly enables the evaluation of inertia at the microscale.

\subsection{Kinematic Links to the Macroscale}  \label{sec:displacementconstraint}
As depicted in Figure \ref{fig:framework}, the macroscopic displacements and deformation gradient and their time derivatives are used to define boundary conditions on the RVE. 
Inserting them into~\eqref{eq:deformationsplit} is only the first step. 
If no additional constraint is considered, the BVP will find an equilibrium where the fluctuations~$\widetilde{\bu}$ oppose the applied displacements which results in zero effective {displacement} of the microstructure. 
{This might not seem obvious at first, however without any imposed constraint, the energetically most favorable position of each node will be its initial configuration, as any deviation from it requires energy. Thus, resulting in a microscopic displacement fluctuation field of $\widetilde{\bu} = \ol{\bu} + \ol{\bH}\bX$, c.f. \eqref{eq:displacementsplit}.}
Based on the principal of kinematic admissibility, described in detail in \citep{Blan:2016:vfag} and applied in dynamic settings e.g. in \citep{Souz:2015:arbm,Roca:2018:acmh}, two kinematic links are chosen here, i.e. 
\begin{align} 
\ol{\bF} = \left\langle \bF \right\rangle \quad \text{and} \quad \ol{\bu} = \left\langle \bu\right\rangle\label{eq:constr} .
\end{align}
The first constraint \eqref{eq:constr}$_{1}$ is well known from quasi-static RVE homogenization frameworks. 
It postulates, that the volume average of the microscopic deformation gradients must equal the macroscopic deformation gradient. 
This is usually enforced by choosing appropriate boundary conditions, e.g., linear displacement or periodic boundary conditions. 
The second constraint \eqref{eq:constr}$_{2}$ is a necessary expansion for the dynamic microscopic problem. 
This link to the macroscopic displacements is essential in order to prevent the RVE from moving arbitrarily in space. 
In quasi-static calculations, fluctuations e.g. of a corner node in the RVE are restricted, which does not influence the results. 
This is, however, not directly possible for the dynamic case without artificially restricting the fluctuations. 
This is due to the fact that the microscopic deformations are already influenced by just moving the RVE. 
Thus, restricting the movement at selected locations will yield different deformation fields. 
Some dynamic homogenization frameworks, e.g. the one in \citep{Pham:2013:tchf}, apply a displacement constraint only on the boundary, which can be a reasonable assumption for dynamic metamaterials. 
There, the boundary lies in the matrix phase, which behaves quasi statically compared to the dynamically active inclusions. 
The framework proposed here does not make any a priori assumptions on which part of the microstructure will be dynamically significant, as this cannot always be determined in advance for arbitrary problems. 
Using the displacement split \eqref{eq:displacementsplit} and the definition of the origin of the local coordinate system as the center of the volume \eqref{eq:origin}, the displacement constraint \eqref{eq:constr}$_{2}$ can be reduced to 
\begin{align}
\left\langle \widetilde{\bu} \right\rangle = \bzero ,\label{eq:dispconst}
\end{align}
which states that the constraint is fulfilled, once the volume average of the fluctuations equals zero. 

\subsection{Algorithmic Treatment of Kinematic Constraints}
To enforce the displacement constraint in~\eqref{eq:dispconst} on the whole RVE domain, we propose to use the method of Lagrange multipliers \citep{bert:1996:colm}. 
Similar applications can be found in \citep{Roca:2018:acmh,Blan:2017:HNS}. 
For this purpose, the mechanical boundary value problem is recast in terms of the principle of minimum potential energy. 
By adding the potential $\Pi^{\lambda}$ associated with the Lagrange multipliers $\Blambda$ and the constraint \eqref{eq:dispconst} to the function of potential energy $\Pi$, one obtains 
\begin{align}
\Pi = \Pi^{\text{int}}  + \Pi^{\text{ext}} + \Pi^{\lambda}, \quad \text{with} \quad  \Pi^{\lambda} = \Blambda\cdot \int_{\B} \widetilde{\bu} \dV .
\end{align}
In the following, only the terms concerning the Lagrange multiplier will be regarded, as the other terms capturing the internal potential energy $\Pi^{\text{int}}$ and the external potential energy $\Pi^{\text{ext}}$ will result in the standard FE formulation \eqref{eq:soe} described above. 
However, note that due to the Lagrange term, the additional degrees of freedom $\Blambda$ appear. 

\subsubsection{Variation}
The potential energy is varied once with respect to the displacement fluctuations $\widetilde{\bu}$ and once with respect to the Lagrange multipliers $\Blambda$, i.e. 
\begin{align}
\delta_{\widetilde{\bu}} \Pi^{\lambda}\,& = \Blambda \cdot  \int_{\B}\delta \widetilde{\bu} \dV\quad\text{and}\label{eq:PIlgr1}\\
\delta_{\Blambda} \Pi^{\lambda} &= \delta \Blambda \cdot  \int_{\B} \widetilde{\bu} \dV .\label{eq:PIlgr2}
\end{align}

\subsubsection{Discretization}
Using $\widetilde{\bu} \approx \bN^{e} \widetilde{\bd}^{e}$ and $\delta \widetilde{\bu} \approx \bN^{e} \delta \widetilde{\bd}^{e}$ as FE approximations, the discretized expressions can be written as
\begin{align}
\delta_{\widetilde{\bu}} \Pi^{\lambda}\,& =\Blambda^{\text{T}} \A_{e}\left[ \int_{\B^{e}}\bN^{e} \dV   \delta \widetilde{\bd}^{e}\right] \quad \text{and}\\
\delta_{\Blambda}  \Pi^{\lambda} &= \delta \Blambda^{\text{T}} \A_{e}\left[  \int_{\B^e}\bN^{e} \dV  \widetilde{\bd}^{e}\right]  .
\end{align}
To obtain the equivalent of the global volume integral in terms of the elements, the assembly operator $\A$ is applied for the respective matrices. 
For better readability, a new element matrix is defined as 
\begin{align}
\bg^{e\text{T}} = \int_{\B^{e}}\bN^{e} \dV   .  \label{eq:gele}
\end{align}
This simplifies the formulations to
\begin{align}
\delta_{\widetilde{\bu}} \Pi^{\lambda}\,& =\Blambda^{\text{T}} \A_{e}\left[\bg^{e{\text{T}}} \delta \widetilde{\bd}^{e}\right] \label{eq:element1} \quad \text{and}\\
\delta_{\Blambda}  \Pi^{\lambda} &= \delta \Blambda^{\text{T}} \A_{e}\left[  \bg^{e{\text{T}}} \widetilde{\bd}^{e}\right]. \label{eq:element2}
\end{align}

\subsubsection{Global Matrix Notation}
To write the whole system of equations as a global problem, the global matrices are defined in terms of the element matrices, i.e. 
\begin{align}
\bG = \A_{e}\bg^{e} ,\quad
\widetilde{\bD} = \U_{e}\widetilde{\bd}^{e}\quad \text{and} \quad
\delta\widetilde{\bD} = \U_{e}\delta\widetilde{\bd}^{e},\label{eq:globalm}
\end{align}
where $\A$ is the afore-mentioned assembly operator and $\U$ a unification operator, as the node displacement fluctuations shared by different elements are not added up, but belong to the same degree of freedom. 
Now the expressions~\eqref{eq:element1} and~\eqref{eq:element2} can be reformulated in global fields as 
\begin{align}
\delta_{\widetilde{\bu}} \Pi^{\lambda}\,& =\Blambda^{\text{T}}\bG^{\text{T}} \delta \widetilde{\bD}=\delta \widetilde{\bD}^{\text{T}}\bG\Blambda\quad \text{and}\label{eq:weak1}\\
\delta_{\Blambda}  \Pi^{\lambda} &= \delta \Blambda^{\text{T}}   \bG^{\text{T}}\widetilde{\bD} = \bzero .\label{eq:weak2}
\end{align}
Since the Lagrange multiplier only appears in~$\Pi^{\lambda}$, no terms result from the variation of $\Pi^{\text{int}}$ and $\Pi^{\text{ext}}$ with respect to $\Blambda$. It follows, that the second expression has to vanish, see~\eqref{eq:weak2}. 

\subsubsection{Linearization}
In order to solve the nonlinear global system of equations, the Newton-Raphson method is utilized. 
For that purpose, we not only need the equations in weak form as in \eqref{eq:weak1} and \eqref{eq:weak2} but also their linearizations. 
They are used to iteratively compute the nodal displacement fluctuations as well as the Lagrange multipliers. 
Here the definition of the \mbox{$\Delta$ operator} is used. When linearizing a function $f(x)=0$ at $\hat{x}$ as Lin$f= f|_{\substack{\hat{x}}} + \Delta f|_{\substack{\hat{x}}}$, then $\Delta f = \frac{\partial f}{\partial x}\big{|}_{\substack{\hat{x}}} \Delta x$. 
Applying this to the weak forms results in
\begin{align}
\text{Lin}\,\delta_{\widetilde{\bu}} \Pi^{\lambda}\,& =\delta \widetilde{\bD}^{\text{T}}\bG \Blambda + \delta \widetilde{\bD}^{\text{T}}\bG\Delta\Blambda \quad\text{and}\\
\text{Lin}\,  \delta_{\Blambda_{1}} \Pi^{\lambda} &= \delta \Blambda^{\text{T}}  \bG^{\text{T}}\widetilde{\bD} + \delta \Blambda^{\text{T}} \bG^{\text{T}}\Delta\widetilde{\bD} = \bzero .
\end{align}

\subsection{Global Discretized Microscopic Problem Including Constraint}
From the linearized variations of $\Pi^{\lambda}$ we define the global residua
\begin{align}
\bR^{\widetilde{\bu}} = -\bG \Blambda  = \A^{e}\bbr^{\widetilde{\bu}^{e}}\quad
\text{with} \quad \bbr^{\widetilde{\bu}^{e}} &= - \bg^{e} \Blambda\quad\text{and} \label{eq:Ru} \\
\bR^{\lambda} =  - \bG^{\text{T}}\widetilde{\bD}= \sum^{e}\bbr^{\lambda^{e}}\quad
\text{with} \quad \bbr^{\lambda^{e}} &= - \bg^{e\text{T}}\widetilde{\bd}^{e}. \label{eq:Rlambda}
\end{align}
Including all linearized variations of $\Pi^{\text{int}}+\Pi^{\text{ext}}+\Pi^{\lambda}$ yields the discrete equation 
\begin{align}
\left[ \begin{array}{c|c}	\delta \widetilde{\bD}^{T}&\delta\Blambda^{\text{T}}\end{array} \right] 
\left[ \begin{array}{c|c}	\widehat{\bK}& \bG\\\hline \bG^{\text{T}} &\bzero\phantom{^{T}} \end{array}\right] 
\left[ \begin{array}{c}	\Delta \widetilde{\bD}\\\hline \Delta \Blambda\end{array}\right]  = 
\left[ \begin{array}{c|c}	\delta \widetilde{\bD}^{\text{T}}&\delta\Blambda^{\text{T}}\end{array} \right] 
\left[ \begin{array}{c}	\widehat{\bR} +\bR^{\widetilde{\bu}}   \\\hline\bR^{\lambda} \end{array} \right] 
\end{align}
as expansion of~\eqref{eq:soe}. 
After including Dirichlet boundary conditions and applying standard arguments of variational calculus, the resulting discrete system of equations reads 
\begin{align}
\left[ \begin{array}{c|c}	\widehat{\bK}& \bG\\\hline \bG^{\text{T}} &\bzero\phantom{^{T}} \end{array}\right] 
\left[ \begin{array}{c}	\Delta \widetilde{\bD}\\\hline \Delta \Blambda\end{array}\right]  = 
\left[ \begin{array}{c}	\widehat{\bR} +\bR^{\widetilde{\bu}}   \\\hline\bR^{\lambda} \end{array}\right]  .\label{eq:lgrsystem}
\end{align}
Note that in contrast to $\bK$, the new tangent stiffness matrix is not necessarily positive definite, which needs to be taken into account when choosing and setting up a solver.  In general, Lagrange multipliers have the disadvantage of adding new degrees of freedom to the system of equations. 
For the presented displacement constraint, only one extra degree of freedom is added for each spacial direction
This is due to the fact that the constraint is applied on the whole RVE which avoids the approximation of the Lagrange multipliers as field variables. 
Thus, for three-dimensional problems,~$\Blambda$ will only {add} three additional degrees of freedom. 
Compared to the displacement fluctuations which are linked to the nodes and which may thus easily reach thousands of degrees of freedom, the number of three additional degrees of freedom over the whole RVE is negligible, making it computationally cheap. 

\subsection{Coupling of the Deformation Gradient}
The constraint related to the deformation gradient \eqref{eq:constr}$_{1}$, can be derived and applied in the same manner as just presented for \eqref{eq:constr}$_{2}$ in the previous section. 
The only change in the final formulation is that the related $\bg^{e\text{T}}_{\langle\bF\rangle}$ matrix needs to be computed as the volume average of the element B-Matrix instead of the shape functions, 
\begin{align}
\bg^{e{\text{T}}}_{\left\langle  \bF \right\rangle}= \int_{\B^{e}}\bB^{e} \dV    .
\end{align}
Applying the constraint regarding the deformation gradient on the volume instead of enforcing it using periodic boundary conditions will lead to minimally invasive boundary conditions enabling e.g. arbitrary damage propagation without artificial restrictions imposed by periodic boundary conditions. 
As shown in \citep{souz:2010:vfol}, such minimally invasive boundary conditions result in a softer constraint compared to periodic boundary conditions. 
{To simplify} the numerical examples in this paper, only the displacement constraint is applied and the constraint related to the deformation gradient is enforced by using periodic boundary conditions.


\FloatBarrier
\section{The Macroscopic Problem \label{sec:macroscale}}
For the solution of the dynamic macroscopic boundary value problem, the associated linearized, discretized balance equations are derived. 
Herein, specific macroscopic tangent moduli appear which are consistently derived for the case where the displacement constraint proposed in the previous section is taken into account. 

\subsection{Macroscopic Boundary Value Problem}

\subsubsection{Macroscopic Equilibrium Equation}
The complete macroscopic balance of linear momentum including inertia is given by 
\begin{align}
\Div\ol{\bP}  + \ol{\bbf} = \bzero.
\end{align}
Applying the Galerkin method with a test function $\delta \ol{\bu}$ on the entire domain $\ol{\B}$ leads to the weak form of linear momentum $\int_{\ol{\B}}\delta\ol{\bu}^{\text{T}}\left( \Div \ol{\bP} + \ol{\bbf}\right)  \dV = 0$. 
By applying  $\Div( {\bP}) \delta {\bu}=  \Div( {\bP}^{\text{T}}\delta {\bu}) - {\bP}\!:\!\Grad \delta {\bu}$ and the Gauss theorem $\int_{{\B}}\Div( {\bP}^{\text{T}}\delta {\bu}) \dV=\int_{\partial{\B}} \delta {\bu} \cdot \bt \,\mathrm{d}A$, the weak form is written as
%
%
\begin{align}
\ol{G} :=
\int_{\ol{\B}}  \delta \ol{\bF}: \ol{\bP} \dV 
+ \int_{\ol{\B}}\delta \ol{\bu}^{\text{T}} \ol{\bbf}^{\rho} \dV 
= 0 .\label{eq:macror}
\end{align}
Herein, zero traction forces are taken into account at the boundary and $\delta\bF = \Grad\delta\bu$. 
Analogous to the microscale, only body forces related to inertia, not gravitation, are considered at the macroscale. 
Thus, the body force vector is set to $\ol{\bbf} \coloneqq \ol{\bbf}^{\rho}= \left\langle \bbf^{\rho}\right\rangle $. 

\subsubsection{Linearization}
To solve the weak form of equilibrium by using the standard Newton-Raphson scheme, the linearized balance of linear momentum is obtained as 
\begin{align}
\text{Lin}\ol{G}  =
\ol{G} + \Delta \ol{G} = 0 \quad\text{with}\quad
\Delta \ol{G}  = \int_{\ol{\B}}  \delta \ol{\bF} :\Delta \ol{\bP}  \dV + \int_{\ol{\B}}\delta \ol{\bu}^{\text{T}} \Delta \ol{\bbf}^{\rho} \dV .
\end{align}
Now the $\Delta$ operator is applied again to $\Delta \ol{\bP}$ and $\Delta \ol{\bbf}^{\rho}$: 
\begin{align}
\Delta \ol{\bP} &= \frac{\partial \ol{\bP}}{\partial\ol{ \bF }}: \Delta \ol{\bF} + \frac{\partial \ol{\bP}}{\partial \ddot{\ol{\bu}}}\cdot \Delta \ddot{\ol{\bu}} \quad \text{and}\\
\Delta \ol{\bbf}^{\rho}&= \frac{\partial \ol{\bbf}^{\rho}}{\partial \ol{\bF}}: \Delta \ol{\bF} + \frac{\partial \ol{\bbf}^{\rho}}{\partial \ddot{\ol{\bu}}}\cdot \Delta \ddot{\ol{\bu}}.
\end{align}
Here an interesting property of the two-scale homogenization framework for dynamics is observed. The macroscopic stress not only depends on the deformation gradient but on the acceleration as well. In turn, the inertia forces can also depend on the deformation gradient in addition to the acceleration. 
We define the resulting four sensitivities as 
\begin{align}
\tangPF\! = \partial_{\ol{\bF}}\ol{\bP},
\quad
\tangPu\! = \partial_{\ddot{\ol{\bu}}}\ol{\bP},
\quad
\tangfF\! = \partial_{\ol{\bF}}\ol{\bbf}^{\rho}
\quad\text{and}\quad
\tangfu\! = \partial_{\ddot{\ol{\bu}}}\ol{\bbf}^{\rho}.\label{eq:tangents}
\end{align}
These moduli  \eqref{eq:tangents} are inserted into the linearized weak form which results in 
\begin{align}
\text{Lin}\ol{G}  = &
\int_{\ol{\B}}  \delta \ol{\bF}: \ol{\bP} \dV 
+ \int_{\ol{\B}}\delta \ol{\bu}^{\text{T}} \ol{\bbf}^{\rho} \dV 
+\int_{\ol{\B}} \delta \ol{\bF}:\tangPF\! : \Delta\ol{\bF} \dV 
\nonumber\\
+&\int_{\ol{\B}} \delta \ol{\bF}:\tangPu\! \cdot \Delta\ddot{\ol{\bu}} \dV 
+ \int_{\ol{\B}}\delta \ol{\bu}^{\text{T}}\tangfF\!: \Delta\ol{\bF} \dV 
+ \int_{\ol{\B}}\delta \ol{\bu}^{\text{T}}\tangfu\!\cdot \Delta\ddot{\ol{\bu}} \dV .
\label{eq:macrolinearization}
\end{align}

\subsubsection{FE Discretization}
The linearization is of the weak form of the balance of linear momentum is now discretized in terms of finite elements. 
First, the linear increment 
\begin{align}
\Delta \ol{G} = &\int_{\B} \delta \ol{\bF}: \tangPF\!: \Delta\ol{\bF} \dV 
+\int_{\B} \delta \ol{\bF}: \ol{\IA}^{P\!,u}\!\cdot \Delta\ddot{\ol{\bu}} \dV\nonumber\\ 
&+ \int_{\B}\delta \ol{\bu}^T \ol{\IA}^{f\!,F}\!: \Delta\ol{\bF} \dV 
+ \int_{\B}\delta \ol{\bu}^T  \ol{\IA}^{f\!,u}\!\cdot \Delta\ddot{\ol{\bu}} \dV ,
\end{align}
is discretized using standard FE formulations. Then, to get rid of the dependence on the time derivatives, the numerical time integration in terms of~\eqref{eq:ti1} is used, which results in 
%
\begin{align}
\Delta \ol{G} = \sum_{e=1}^{n_{\text{ele}}} \delta\ol{ d}^e_P&\left(  \int_{\B^e} \ol{B}^e_{ijP} \ol{\IA}^{P\!,F}_{ijmn} \ol{B}^e_{mnQ}  
+ \frac{\ol{\alpha}_1}{\Delta t^2}   \ol{B}^e_{ijP} \ol{\IA}^{P\!,u}_{ijk}\ol{N}^e_{Qk}\right. \nonumber\\
&\left.	+  \ol{N}^e_{Pi} \ol{\IA}^{f\!,F}_{imn}\ol{ B}^e_{mnQ}  
+ \frac{\ol{\alpha}_1}{\Delta t^2} \ol{N}^e_{Pi}  \ol{\IA}^{f\!,u}_{ik}\ol{N}^e_{Qk} \dV  \right)  \Delta{\ol{d}}_Q^e .
\end{align}
Herein, the matrix representation of the moduli {in} index notation has been used. {Lowercase indices refer the spacial dimension $n_{\text{dm}}$, whereas uppercase indices to the total degrees of freedom of a element $n_{\text{edf}}$.}
Again, standard element B-matrix $\ol{B}^e$ and shape function matrix $\ol{N}^e$ are considered. 
By extracting the nodal virtual and incremental displacements, this yields the definition of the full macroscopic element tangent stiffness matrix 
\begin{align}
\widehat{\ol{k}}_{PQ}^e = \int_{\B^e}&\left(  \ol{B}^e_{ijP} \ol{\IA}^{P\!,F}_{ijmn} \ol{B}^e_{mnQ}  
+ \frac{\ol{\alpha}_1}{ \Delta t^2}  \ol{ B}^e_{ijP} \ol{\IA}^{P\!,u}_{ijk}\ol{N}^e_{Qk} \right.\nonumber\\
&\left. +  \ol{N}^e_{Pi} \ol{\IA}^{f\!,F}_{imn} \ol{B}^e_{mnQ}  
+ \frac{\ol{\alpha}_1}{ \Delta t^2} \ol{N}^e_{Pi}  \ol{\IA}^{f\!,u}_{ik}\ol{N}^e_{Qk} \right) \dV . \label{eq:mackele}
\end{align}
Now the remaining part of the linearization~\eqref{eq:macrolinearization}, the residuum $\ol{R}$, is discretized as 
\begin{align}
\widehat{\ol{R}} = \sum_{e=1}^{n_{\text{ele}}} \left( \delta \ol{d}_P \int_{\B^e} \ol{B}^e_{ijP} \ol{P}_{ij} \dV 
+ \delta \ol{d}_P \int_{\B^e}\ol{N}^e_{iP}\ol{ f}^{\rho}_{i}   \dV   \right) .
\end{align}
By extracting the nodal virtual displacements, the element residuum is identified as 
\begin{align}
\widehat{\ol{r}}_{P}^e = \int_{\B^e}& \left(   \ol{ B}^e_{ijP}\ol{ P}_{ij}
+ \ol{N}^e_{Pi} \ol{f}^{\rho}_{i}   \right)  \dV  , \label{eq:macrele}
\end{align}
where again matrix representation and index notation is used. 

\subsection{Consistent Macroscopic Tangent Moduli}

For the dynamic homogenization framework, four macroscopic tangent moduli \eqref{eq:tangents} need to be determined. 
To obtain the sought-after moduli in closed form, we start with taking the derivative of the incremental linearized weak form of linear momentum at the microscale with respect to the two relevant macroscopic quantities, the deformation gradient $\ol{\bF}$ and the acceleration $\ddot{\ol{\bu}}$. 
Then we derive the moduli by considering the microscopic problem in its equilibrium state. 

\subsubsection{Incremental Weak Forms Including Displacement Constraint \label{ssec:wlm}} 
As it will be shown later, the derivatives of the microscopic fluctuations with respect to the macroscopic deformation gradient and the acceleration will be required. 
For their calculation, the incremental linearized weak forms have to be derived with respect to these two quantities. 
In order to account for the proposed displacement constraint, the associated increments of the weak form of linear momentum and of the Lagrange multiplier potential {with respect to the microscopic displacement fluctuations as well as the Lagrange multipliers}, i.e.~\eqref{eq:PIlgr1} and~\eqref{eq:PIlgr2}, are identified as 
\begin{align}
\Delta G^{\widetilde{u}} &= \int_{\B} \delta F_{ij} \IA_{ijmn }\Delta F_{mn} \,\mathrm{d}V + \int_{\B} \delta \widetilde{u}_{i}\rho_{0}\Delta \ddot{u}_{i}\,\mathrm{d}V + \Delta \lambda_{i} \int_{\B}\delta \widetilde{u}_{i} \dV\label{eq:ilwlm}\quad\text{and}\\
\Delta G^{\lambda} &= \delta \lambda_{i}  \int_{\B} \Delta \widetilde{u}_{i} \dV .\label{eq:wlmlgr}
\end{align}
Using the decompositions~\eqref{eq:displacementsplit} and~\eqref{eq:Fsplit}, equation \eqref{eq:ilwlm} can be reformulated as 
\begin{align}
\Delta G^{\widetilde{u}} = &\int_{\B} \delta F_{ij} \IA_{ijmn }\Delta \ol{F}_{mn} \dV +
\int_{\B} \delta F_{ij} \IA_{ijmn }\Delta \widetilde{H}_{mn} \dV 
+\int_{\B} \delta u_{i}\rho_{0}\Delta  \ddot{\ol{u}}_{i}\dV   \nonumber\\
&+
\int_{\B} \delta u_{i}\rho_{0}\Delta  \ddot{\ol{F}}_{ij}X_{j}\dV  +
\int_{\B} \delta u_{i}\rho_{0}\Delta  \ddot{\widetilde{u}}_{i}\dV +
\Delta \lambda_{i} \int_{\B}\delta \widetilde{u}_{i} \dV .\label{eq:wlm}
\end{align}

\subsubsection{Derivatives of Incremental Weak Forms}

By taking the derivatives of the increments \eqref{eq:wlmlgr} and \eqref{eq:wlm} in the equilibrium state $\Delta G = 0$, a closed form formulation of the tangent moduli will be obtained later. 
Thus, the associated derivatives are computed in the following. 

{\bfseries Derivative with Respect to $\ol{\bF}$:}
Taking the derivatives of \eqref{eq:wlmlgr} and \eqref{eq:wlm} with respect to the macroscopic deformation gradient, while considering \eqref{eq:ti1} results in 
\begin{align}
0_{kl} = &\int_{\B} \delta F_{ij} \IA_{ijkl} \dV +
\int_{\B} \delta F_{ij} \IA_{ijmn }\frac{\partial  \widetilde{H}_{mn}}{\partial \ol{F}_{kl}}\dV  +
\frac{\ol{\alpha}_1}{ \Delta t^{2}}\int_{\B} \delta u_{k}\rho_{0} X_{l}\dV \nonumber\\ 
&  +\frac{\alpha_1}{ \Delta t^{2}}\int_{\B} \delta u_{i}\rho_{0}  \frac{\partial  \widetilde{u}_{i}}{\partial \ol{F}_{kl}}         \dV  + 
 \frac{\partial \lambda_{i}}{\partial \ol{F}_{kl}}  \int_{\B}\delta  \widetilde{u}_{i} \dV\quad\text{and}\\
0_{kl} = & \delta \lambda_{i} \int_{\B} \frac{\partial  \widetilde{u}_{i}}{\partial \ol{F}_{kl}} \dV .
\end{align}
Using standard FE discretization yields
\begin{align}
0_{kl} = &\sum_{e=1}^{n_{\text{ele}}} \delta \widetilde{d}_{P}^{e} \left(  \int_{\B^{e}} B^{e}_{ijP} \IA_{ijkl} \dV +
\int_{\B^{e}} B^{e}_{ijP}\IA_{ijmn }B^{e}_{mnQ}\dV \frac{\partial  \widetilde{d}^{e}_{Q}}{\partial \ol{F}_{kl}} 
+ \frac{\ol{\alpha}_1}{\Delta t^{2}}\int_{\B^{e}}N^{e}_{Pk} \rho_{0} X_{l}\dV 
\right. \nonumber\\ 
&+\left. 
\frac{\alpha_1}{\Delta t^{2}}\int_{\B^{e}} N^{e}_{Pi}\rho_{0}  N^{e}_{Qi}\dV \frac{\partial  \widetilde{d}^{e}_{Q}}{\partial \ol{F}_{kl}}  +
 \int_{\B} N^{e}_{Pi} \dV \frac{\partial \lambda_{i}}{\partial \ol{F}_{kl}} 
\right)  \quad\text{and}\\
0_{kl} = &\sum_{e=1}^{n_{\text{ele}}} \delta \lambda_{i}  \left( \int_{\B}  N^{e}_{Pi}  \dV  \frac{\partial  \widetilde{d}^e_{P}}{\partial \ol{F}_{kl}} \right) .
\end{align}
Rewriting this in global notation using the abbreviations defined in Appendix \ref{sec:appendixA} Table \ref{tab:m_overview} leads to the expressions 
\begin{align}
\bzero &= \bL + \frac{\ol{\alpha}_1}{\Delta t^{2}} \bZ + \left(\bK+
\frac{\alpha_1}{\Delta t^{2}} \bM\right)  \frac{\partial  \widetilde{\bD}}{\partial \ol{\bF}}
 + \bG \frac{\partial  \Blambda}{\partial \ol{\bF}} \quad \text{and}\\
 \bzero &= \bG^\text{T} \frac{\partial  \widetilde{\bD}}{\partial \ol{\bF}} .
\end{align}
By combining the nodal fluctuations and the Lagrange multipliers into one column matrix~$\bD^*$, the two equations can be written as one system of equations $\bK^*\partial_{\ol{\bF}}\bD^*=-\bL^{\ol{*}}$ with the matrices 
\begin{align}
\bD^{*\text{T}}=&
\left[ \begin{array}{c|c}	 \widetilde{\bD}^{\text{T}} &	\Blambda^{\text{T}} \end{array}\right],\label{eq:def1}\\
\bL^{\ol{*}^\text{T}}=& 
\left[ \begin{array}{c|c}	 \bL^{\text{T}}  + \frac{\ol{\alpha}_1}{\Delta t^{2}} \bZ^{\text{T}}   & 	\bzero  \end{array}\right] \quad \text{and}\label{eq:def2}\\
\bK^{{*}}=& 
\left[ \begin{array}{c|c}	 \bK + \frac{{\alpha_1}}{\Delta t^{2}} \bM   & 	\bG \\ \hline	 \bG^{\text{T}} & \bzero \end{array}\right]  .\label{eq:def3}
\end{align}
Then, the required derivative can be computed from 
\begin{align}
\frac{\partial  {\bD}^*}{\partial \ol{\bF}}
 = -\bK^{*^{-1}}  \bL^{\ol{*}} . \label{eq:dDdF}
\end{align}
Note that $\bK^*$ is the microscopic tangent stiffness matrix in~\eqref{eq:lgrsystem}, which is already available from solving the microscopic boundary value problem. 

{\bfseries Derivative with Respect to $\ddot{\ol{\bu}}$:}
Analogously, the derivative of~\eqref{eq:wlm} with respect to $\ddot{\ol{\bu}}$ can be obtained by applying~\eqref{eq:ti2}, i.e. one obtains 
\begin{align}
0_k = &
\frac{\Delta t^2}{\alpha_1}\int_{\B} \delta F_{ij} \IA_{ijmn }\frac{\partial \ddot{\widetilde{H}}_{mn}}{\partial \ddot{\ol{u}}_{k}}\dV +
 \int_{\B} \delta u_{k}\rho_{0} \dV + 
\int_{\B} \delta u_{i}\rho_{0}  \frac{\partial  \ddot{\widetilde{u}}_{i}}{\partial \ddot{\ol{u}}_{k}}\dV   + 
\frac{\partial \lambda_{i}}{\partial \ddot{\ol{u}}_{k}}  \int_{\B}\delta  \widetilde{u}_{i} \dV\\
0_{k} = & \delta \lambda_{i} \int_{\B} \frac{\partial  \widetilde{u}_{i}}{\partial\ddot{\ol{u}}_{k}} \dV . 
\end{align}
Standard FE discretization using matrix representation and index notation yields 
\begin{align}
0_k = &\sum_{e=1}^{n_{\text{ele}}} \delta \widetilde{d}_{P}^{e} \left( 
\int_{\B^{e}} B^{e}_{ijP}\IA_{ijmn }B^{e}_{mnQ}\dV \frac{\partial  {\widetilde{d}}^{e}_{Q}}{\partial \ddot{\ol{u}}_{k}} + 
\int_{\B^{e}} N_{Pk}\rho_{0} \dV \right. \nonumber\\ 
&\left. +  
\frac{\alpha_1}{\Delta t^{2}}\int_{\B^{e}} N^{e}_{Pi}\rho_{0}  N^{e}_{Qi}\dV \frac{\partial  {\widetilde{d}}^{e}_{Q}}{\partial \ddot{\ol{u}}_{k}}   +
\int_{\B} N^{e}_{Pi} \dV \frac{\partial \lambda_{i}}{\partial \ddot{\ol{u}}_{k}} 
\right)  \quad\text{and}\\
0_{k} = &\sum_{e=1}^{n_{\text{ele}}} \delta \lambda_{i}  \left( \int_{\B}  N^{e}_{Pi}  \dV  \frac{\partial  \widetilde{d}^e_{P}}{\partial \ddot{\ol{u}}_{k}} \right) .
\end{align}
Again, the two equations are combined by {joining} the displacements and the Lagrange multipliers {into a single matrix, c.f. \eqref{eq:def1}}. 
By solving the resulting system of equations with respect to the required derivatives $\partial_{\ddot{\ol{\bu}}}  \bD^{*}$, we obtain 
\begin{align}
\frac{\partial  {\bD}^*}{\partial \ddot{\ol{\bu}}}
= -\bK^{*^{-1}}  \bW^* . \label{eq:dDdu}
\end{align}
Herein, the definitions~\eqref{eq:def1}-\eqref{eq:def3}, the abbreviations defined in Appendix~\ref{sec:appendixA} Table~\ref{tab:m_overview}, as well as $\bW^{*\text{T}}=\left[\begin{array}{c|c} \bW^{\text{T}}&\bzero  \end{array}\right]$ are used. 

\subsubsection{Derivation of Tangent Moduli \label{ssec:derivation}}
In this subsection the four moduli will be derived by inserting the derivatives computed in the last subsection. 
Note that all moduli are only consistent for a microscopic equilibrium state. 
Thus, quadratic convergence of the macroscopic Newton-Raphson iteration is only ensured if the microscopic boundary value problem is solved for each macroscopic iteration step. 
After the last microscopic iteration, the consistent tangent moduli can be computed. 

{\bfseries Derivation of $\tangPF$:} 
To derive the sensitivity of the macroscopic stresses with respect to the macroscopic deformation gradient, the derivative is rewritten using the definition of the macroscopic stresses in terms of the microscopic fields, i.e. 
\begin{align}
\tangPF_{ijmn} &=  \dfrac{\partial \ol{P}_{ij}}{\partial \ol{F}_{mn}} = \dfrac{\partial  \left\langle P_{ij} +\rho_{0}\ddot{u}_{i} X_{j} \right\rangle }{\partial \ol{F}_{mn}} .
\end{align}
Using the chain rule $\frac{\partial \bP(\bF)}{\partial \ol{\bF}} =   \frac{\partial \bP}{\partial {\bF}} :   \frac{\partial \bF}{\partial \ol{\bF}}$, applying \eqref{eq:displacementsplit}, \eqref{eq:Fsplit}, \eqref{eq:ti1} and inserting FE discretization, the equation can be written as 
\begin{align}
\tangPF_{ijmn} = &\sum_{e=1}^{n_{\text{ele}}} \left(   \frac{1}{V}\int_{\B^{e}}\right.\IA_{ijmn} \dV
+ \frac{\ol{\alpha}_1}{\Delta t^{2}}\frac{1}{V} \int_{\B^{e}}\rho_{0}\delta_{im}X_{j}X_{n} \dV  \notag\\
&+ \frac{1}{V} \int_{\B^{e}}\IA_{ijkl}B^{e}_{klP}\dV\frac{\partial \widetilde{d}_{P}^{e}}{\partial \ol{F}_{mn}}   
+ \left.\frac{\alpha_1}{\Delta t^{2}}\frac{1}{V} \int_{\B^{e}}\rho_{0} N^{e}_{Pi}X_{j} \dV \frac{\partial  \widetilde{d}^{e}_{P}}{\partial \ol{F}_{mn}}\right) .\label{eq:result1}
\end{align}
By using the global abbreviations defined in Appendix \ref{sec:appendixA} Table \ref{tab:m_overview}, Table \ref{tab:m*_overview} and inserting~\eqref{eq:dDdF}, the closed form result is obtained as 
\begin{align}
\tangPF &= \left\langle \IA  + \frac{1}{\ol{\beta}\Delta t^{2}} \IY \right\rangle - \frac{1}{V} \bL^{{*}^\text{T}} \bK^{*^{-1}} \bL^{\ol{*}} .\label{eq:tangPF}
\end{align}
Note that this result has already been {presented} in \citep{tams:2018:fstd} for a special scenario of dynamic homogenization, which did not include macroscopic acceleration and the displacement constraint. 

{\bfseries Derivation of $\tangPu$:} 
The derivation of the sensitivity of the macroscopic stresses with respect to the macroscopic accelerations is analogous to that of $\tangPF\!$. 
First the derivative is rewritten using the definition of the macroscopic stresses in terms of the microscopic fields as 
\begin{align}
\tangPu_{ijk} &=  \dfrac{\partial \ol{P}_{ij}}{\partial \ddot{\ol{u}}_{k}} = \dfrac{\partial  \left\langle P_{ij} +\rho_{0}\ddot{u}_{i} X_{j} \right\rangle }{\partial \ddot{\ol{u}}_{k}} .
\end{align}
Then using  the chain rule, \eqref{eq:displacementsplit}, \eqref{eq:Fsplit}, \eqref{eq:ti1} and FE discretization, the equation reads 
\begin{align}
\tangPu_{ijk} = &\sum_{e=1}^{n_{\text{ele}}}\left(  
\frac{1}{V}\int_{\B^{e}}\rho_{0}\delta_{ik}X_{j}\dV +
\frac{1}{V}\int_{\B^{e}} \IA_{ijmn}B^{e}_{mnP}\dV\frac{\partial \widetilde{d}^{e}_{P}}{\partial \ddot{\ol{u}}_{k}}\right.\nonumber\\
&\left.+\frac{\alpha_1}{\Delta t^{2}}\frac{1}{V}\int_{\B^{e}}\rho_{0}X_{j} N^{e}_{Pi}\dV\frac{\partial  {\widetilde{d}}^{e}_{P}}{\partial \ddot{\ol{u}}_{k}}\right) .\label{eq:Au3}
\end{align}
Finally, using the global abbreviations in Appendix \ref{sec:appendixA} Table \ref{tab:m_overview}, Table \ref{tab:m*_overview} and inserting \eqref{eq:dDdu}, the modulus is obtained as 
\begin{align}
\tangPu &= \left\langle \bV \right\rangle -
 \frac{1}{V} \bL^{{*}^\text{T}} \bK^{*^{-1}}\bW^{{*}}
 .\label{eq:tangPu}
\end{align}

\begin{figure}[t]
	\begin{algorithm}[H]
		\linespread{1.2}\selectfont
		\DontPrintSemicolon
		\SetInd{0.5em}{0.8em}
		\uline{\bfseries{Macroscopic problem}}\;
		\Loop{over all macroscopic elements}{
			\Loop{over element Gauss points}{
				\uline{\bfseries{Microscopic problem}}\;
				\Input{$\ol{\bF},\ol{\bu},{\ddot{\ol{\bF}}},{\ddot{\ol{\bu}}}$}
				\bc{$\bx = {\ol{\bu}} + \ol{\bF}\bX+\widetilde{\bu}$ with ${\left\langle \widetilde{\bu} \right\rangle  = \bzero}$ on $\B$ \eq*[r]{\eqref{eq:deformationsplit},\eqref{eq:dispconst}}}
				\Loop{micro-iteration until $\left|\Delta\bD^{*}\right| < {tol}$}{
					\Compute{global microscopic fields:\\
						$\quad\quad\quad\quad~\widehat{\bK},\widehat{\bR},{\bR^{\widetilde{u}}},{\bR^{\lambda}},{\bG}$} \eq*[r]{\eqref{eq:mickele},\eqref{eq:micrele}, \eqref{eq:Ru},\eqref{eq:Rlambda},\eqref{eq:gele}}
					\Solve{$\bK^*\Delta\bD^* = \bR^*$} \eq*[r]{\eqref{eq:lgrsystem}}  
					\Update{$\bD^{*} \Leftarrow \bD^{*} + \Delta\bD^{*}$}\;
				}
				\Compute{homogenized fields and moduli:\\
					$\quad\quad\quad\quad~\ol{\bP},{\ol{\bbf}^{\rho}},\tangPF\!\!,{\tangPu\!\!},{\tangfF\!\!},{\tangfu}$} 	\eq*[r]{\eqref{eq:macroP},\eqref{eq:macrof},\eqref{eq:tangPF},\eqref{eq:tangPu},\eqref{eq:tangfF},\eqref{eq:tangfu}}
			}
			\Compute{element matrices $\widehat{\ol{\bk}}^e\!\!$ and $\widehat{\ol{\bbr}}^e\!\!$, using Gauss integration}
			\eq*[r]{\eqref{eq:mackele},\eqref{eq:macrele}}	
		}
		\Solve{$\widehat{\ol{\bK}}\Delta\ol{\bD} = \widehat{\ol{\bR}}$} 
	\end{algorithm} 
	\caption{Algorithm for single macroscopic iteration of the dynamic FE$^2$ framework with respective equation references. It should be noted that the overall structure of the standard FE procedure does not change, only some additional fields need to be computed. Furthermore, for the implementation of the microscopic problem, the macroscopic displacments $\ol{\bu}$ may be omitted from the code. It is the second derivative $\ddot{\ol{\bu}}$, computed in the macroscopic problem, which influences the microscopic results.}
	\label{alg:framework}
\end{figure}

{\bfseries Derivation of $\tangfF$:} 
The derivation of the sensitivity of the macroscopic inertia with respect to the macroscopic deformation gradient is again similar to that of $\tangPF\!$. 
First, the derivative is rewritten as 
\begin{align}
\tangfF_{imn} &=  \dfrac{\partial \ol{f}^{\rho}_{i}}{\partial \ol{F}_{mn}} = \dfrac{\partial  \left\langle \rho_{0}\ddot{u}_{i} \right\rangle }{\partial \ol{F}_{mn}} 
\end{align}
and by using \eqref{eq:displacementsplit}, \eqref{eq:Fsplit}, \eqref{eq:ti1} and FE discretization, the equation reads 
\begin{alignat}{2}
\tangfF_{imn} = \quad \sum_{e=1}^{n_{\text{ele}}} \left( 
\frac{\ol{\alpha}_1}{ \Delta t^{2}}\frac{1}{V}\int_{\B^{e}}\rho_{0}\delta_{im}X_{n} \dV
+ \frac{\alpha_1}{\Delta t^{2}}\frac{1}{V}\int_{\B^{e}}\rho_{0} N^{e}_{Pi} \dV \frac{\partial  \widetilde{d}^{e}_{P}}{\partial\ol{F}_{mn}}
\right) .
\end{alignat}
Using the global abbreviations in Appendix \ref{sec:appendixA} Table \ref{tab:m_overview}, Table \ref{tab:m*_overview}  and plugging in \eqref{eq:dDdF}, the modulus is  identified as 
\begin{align}
\tangfF &=  \frac{\ol{\alpha}_1}{\Delta t^{2}}  \left\langle \bV^{\text{T}} \right\rangle- \frac{1}{V} \frac{\alpha_1}{\Delta t^{2}} \bW^{{*}^\text{T}} \bK^{*^{-1}} \bL^{\ol{*}}.\label{eq:tangfF}
\end{align}

{\bfseries Derivation of $\tangfu$:} 
Analogously, the derivative is rewritten as 
\begin{align}
\tangfu_{ik} &=  \dfrac{\partial \ol{f}^{\rho}_{i}}{\partial \ddot{\ol{u}}_{k}} = \dfrac{\partial  \left\langle \rho_{0}\ddot{u}_{i} \right\rangle }{\partial \ddot{\ol{u}}_{k}}.
\end{align}
Then, using~\eqref{eq:displacementsplit} and FE discretization, the expression becomes 
\begin{alignat}{2}
\tangfu_{ik} = \quad \sum_{e=1}^{n_{\text{ele}}} \left( 
\frac{1}{V}\int_{\B^{e}}\rho_{0}\delta_{ik} \dV
+ \frac{1}{V} \frac{\alpha_1}{\Delta t^{2}}\int_{\B^{e}}\rho_{0} N^{e}_{Pi} \dV \frac{\partial  {\widetilde{d}}^{e}_{P}}{\partial\ddot{\ol{u}}_{k}}
\right) .
\end{alignat}
Taking into account the global abbreviations in Appendix \ref{sec:appendixA} Table \ref{tab:m_overview}, Table \ref{tab:m*_overview}  and inserting \eqref{eq:dDdu}, the modulus is derived as 
\begin{align}
\tangfu &=   \left\langle \Brho_{0}\right\rangle -  \frac{1}{V} \frac{\alpha_1}{\Delta t^{2}} \bW^{*^{\text{T}}} \bK^{*^{-1}}\bW^* .\label{eq:tangfu}
\end{align}
Note that if $\ol{\alpha}_1$ and $\alpha_1$ are equal to zero, which would be equivalent to a quasi-static calculation, the first tangent moduli in~\eqref{eq:tangPF} take the same form as e.g. found in \citep{Mieh:1999:cmmt}. 
Here, the {closed form} moduli \eqref{eq:tangPF},\eqref{eq:tangPu}, \eqref{eq:tangfF} and \eqref{eq:tangfu} extend this consistently to the dynamic regime. 
An overview over the algorithm of the proposed framework is presented in Figure \ref{alg:framework}.


\FloatBarrier
\section{Numerical Examples \label{sec:numericalexamples}}
This section presents numerical examples as a proof of concept, as well as an initial analysis of different RVE choices. 
As it turns out, for dynamic homogenization the definition of RVE is even more complex than for quasistatic cases. 
Single-scale comparisons are calculated to asses the reliability of the homogenization framework. 
First, a rather arbitrary example is shown to analyze the macroscopic Newton iteration and demonstrate the quadratically converging algorithm, which is based on the tangent moduli derived in Section~\ref{sec:macroscale}. 
Then the concept of a unit cell as RVE is analyzed under dynamic conditions. 
Finally, a comparison of two different displacement constraints, including the proposed one, is presented. 
All numerical examples make use of the Newmark scheme with the parameters $\gamma = 0.5$ and $\beta = 0.25$, resulting in an unconditionally stable algorithm. 
For more details on time integration methods in the context of nonlinear FE methods see e.g. \citep{wrig:2008:nfem}.\par
\begin{figure}[b]
	\centering
	 
		\tikzset{external/export next=false}
		\begin{tikzpicture}  
		\def\colorA{white}; 
		\def\colorB{gray};  
		\def\colorC{black}; 
		
		\foreach \x in {0,2,4} {
		    \draw [fill=colorA!50,colorA!50] (3+\x,0) rectangle (4+\x,1);
		    \draw [fill=colorB!50,colorB!50] (4+\x,0) rectangle (5+\x,1);};
		\draw (0.5,1)--(0,1)--(0,0)--(0.5,0.);
		\draw[dashed] (0.5,0)--(2.5,0);
		\draw[dashed] (0.5,1)--(2.5,1);
		\draw (2.5,0)--(9.5,0);
		\draw (2.5,1)--(9.5,1);
		\draw[dashed] (9.5,0)--(11.5,0);
		\draw[dashed] (9.5,1)--(11.5,1);
		\draw (11.5,1)--(12,1)--(12,0)--(11.5,0.);
		\foreach \x in {0,...,6}
			\draw (\x+3,0)--(\x+3,1);		
		\foreach \x in {9,...,23}
			\draw (\x*0.25+3,0)--(\x*0.25+3,1);		
		\foreach \x in {0,...,5}
			\draw [line width=.3mm,\colorC] (-0.2,\x*.2-0.1)--(0,\x*.2);		
		\draw[line width=.3mm,\colorC]  (0,-0.2)--(0,1.2);
			
		\def\D{-0.7}; 
		\draw (-0.2,\D)--(12.2,\D)node[right]{\footnotesize  $\displaystyle  {L}$};
		\def\Px{0.0} 
		\def\Py{\D} 
			\draw (\Px,\Py-0.1)--(\Px,\Py+0.1);
			\draw (\Px-0.075,\Py-0.075)--(\Px+0.075,\Py+0.075);
		\def\Px{12.0} 
		\def\Py{\D} 
		\draw (\Px,\Py-0.1)--(\Px,\Py+0.1);
		\draw (\Px-0.075,\Py-0.075)--(\Px+0.075,\Py+0.075);
		
		\draw (4-0.2,-.3)--(5.2,-.3)node[right]{\footnotesize  $\displaystyle l_{\text{M}}$};
		
		\def\Px{4.0} 
		\def\Py{-.3} 
		\draw (\Px,\Py-0.1)--(\Px,\Py+0.1);
		\draw (\Px-0.075,\Py-0.075)--(\Px+0.075,\Py+0.075);
		\def\Px{5.0} 
		\def\Py{-.3} 
		\draw (\Px,\Py-0.1)--(\Px,\Py+0.1);
		\draw (\Px-0.075,\Py-0.075)--(\Px+0.075,\Py+0.075);

		\draw (7.25-0.2,-.3)--(7.5+.2,-.3)node[right]{\footnotesize  $\displaystyle l_{\text{E}}$};
		\def\Px{7.25} 
		\def\Py{-.3} 
		\draw (\Px,\Py-0.1)--(\Px,\Py+0.1);
		\draw (\Px-0.075,\Py-0.075)--(\Px+0.075,\Py+0.075);
		\def\Px{7.5} 
		\def\Py{-.3} 
		\draw (\Px,\Py-0.1)--(\Px,\Py+0.1);
		\draw (\Px-0.075,\Py-0.075)--(\Px+0.075,\Py+0.075);
			\draw[-latex,line width=1mm, \colorC]   (12.7,.5) to (12,.5)   ;
			\node at (13.2,.5) {\footnotesize  $\displaystyle {\ol{u}(t)}$};

			\node at (3.5,.5) {\footnotesize  $\displaystyle E_1, \rho_1$};					
			\node at (4.5,.5) {\footnotesize  $\displaystyle E_2, \rho_2$};								
			\node at (0,1.5)[right]{Direct Numerical Simulation (DNS)};												
			\node at (-0.6,1.5)[right]{(a)};					
		\end{tikzpicture}
		
			\tikzset{external/export next=false}
			\begin{tikzpicture}  
			\def\colorA{white};
			\def\colorB{gray};
			\def\colorC{black};
			
			\draw (0,0)--(0.5,0.);
			\draw[dashed] (0.5,0)--(2.5,0);
			\draw (2.5,0)--(9.5,0);
			\draw[dashed] (9.5,0)--(11.5,0);
			\draw(12,0)--(11.5,0.);
			\foreach \x in {1,...,3}			
			\draw (\x*3,-0.2)--(\x*3,0.2);
			\draw[line width=.3mm,\colorC] (12,-0.2)--(12,0.2);
			
			\foreach \x in {-1,...,1}
			\draw [line width=.3mm,\colorC] (-0.2,\x*.2-0.1)--(0,\x*.2);		
			
			\draw[line width=.3mm,\colorC]  (0,-0.4)--(0,0.4);
			
			\draw (3-0.2,-.5)--(6+0.2,-.5)node[pos=.5,below]{\footnotesize  $\displaystyle l_{\ol{\text{E}}}$};
			
			\def\Px{3} 
			\def\Py{-.5} 
			\draw (\Px,\Py-0.1)--(\Px,\Py+0.1);
			\draw (\Px-0.075,\Py-0.075)--(\Px+0.075,\Py+0.075);
			
			\def\Px{6} 
			\def\Py{-.5} 
			\draw (\Px,\Py-0.1)--(\Px,\Py+0.1);
			\draw (\Px-0.075,\Py-0.075)--(\Px+0.075,\Py+0.075);
			\foreach \x in {3.63, 5.37, 6.63, 8.37}{
				\draw (\x+0.2,-.2)--(\x-0.2,0.2);
				\draw (\x-0.2,-.2)--(\x+0.2,0.2);};
			
			\def\l{4.8} 
			\def\h{0.3} 
			\def\d{0.5}
			\def\D{0.3}
			
			\foreach \x in {8.37}{		
			\draw [fill=colorB!50,colorB!50] (\x-\l/4,-\h-\d) rectangle (\x-\l/8,-\d);
			\draw [fill=colorB!50,colorB!50] (\x+\l/4,-\h-\d) rectangle (\x+\l/8,-\d);
			\draw [fill=colorA!50,colorA!50] (\x-\l/8,-\h-\d) rectangle (\x+\l/8,-\d);		
				
				\draw (\x-0.2,-0.1)--(\x-\l/4,-\d);
				\draw (\x+.2,-0.1)--(\x+\l/4,-\d);
		    \draw (\x+\l/4,-\d)--(\x-\l/4,-\d);
			    \draw (\x+\l/4,-\h-\d)--(\x-\l/4,-\h-\d);
			
		   \foreach \y in {0,1,...,8}
	       \draw (\y*\l/16+\x-\l/4,-\d)--(\y*\l/16+\x-\l/4,-\h-\d);	
			       
			    \node at (\x+\l/4+1.2,-\d-\h/2) {\footnotesize example RVE};		
			    \node at (\x,0.5) {\footnotesize Integration Point};

			    \draw (\x-\l/16-0.2,-\d-\h-\D)--(\x+.2,-\d-\h-\D)node[right]{\footnotesize  $\displaystyle l_{\text{E}}$};
			    \def\Px{\x-\l/16} 
			    \def\Py{-\d-\h-\D} 
			    \draw (\Px,\Py-0.1)--(\Px,\Py+0.1);
			    \draw (\Px-0.075,\Py-0.075)--(\Px+0.075,\Py+0.075);
			    \def\Px{\x} 
			    \def\Py{-\d-\h-\D} 
			    \draw (\Px,\Py-0.1)--(\Px,\Py+0.1);
			    \draw (\Px-0.075,\Py-0.075)--(\Px+0.075,\Py+0.075);

			};
			\draw[-latex,line width=1mm, \colorC]   (12.7,.0) to (12,.0)   ;
			\node at (13.1,0) {\footnotesize  $\displaystyle {\ol{u}(t)}$};						
			\node at (0,0.7)[right]{Multiscale Simulation};																
			\node at (-0.6,0.7)[right]{(b)};					
			\end{tikzpicture}
	\caption{Illustration of the numerical calculations including (a) a 1D single-scale FE Model and (b) the macroscopic model and the RVE of the FE$^2$ approach. 
		\label{fig:DNS_FE2}}
\end{figure}
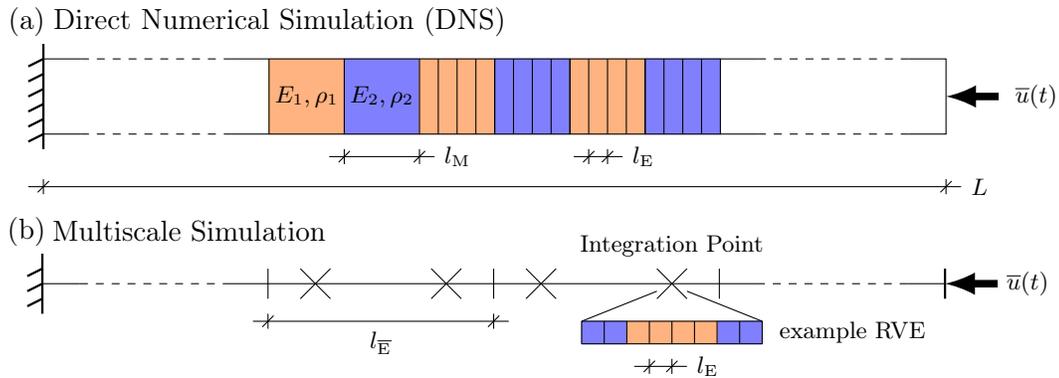
A one-dimensional model of a layered structure with the total length of ${L}$ is investigated. 
The studied heterogeneous material consists of two alternating phases, a soft and light phase, and a stiff and heavy phase. 
Each phase has a length of $l_{\text{M}}$, a Young's modulus $E_1$ and $E_2$, and a density $\rho_1$ and $\rho_2$, respectively. 
All calculations are run using $E_1=2\cdot 10^{3}~\frac{\text{N}}{\text{mm}^{2}}$, $E_2=2\cdot 10^{5}~\frac{\text{N}}{\text{mm}^{2}}$, $\rho_1=1\cdot 10^{3}~\frac{\text{kg}}{\text{m}^{3}}$ and $\rho_2=1\cdot 10^{5}~\frac{\text{kg}}{\text{m}^{3}}$. 
The Poisson's ratio is chosen to be negligible, i.e. $\nu = 10^{-6}$, to enable a \mbox{quasi-1D} investigation. 
The left boundary is fixed, on the right end an impact load is applied in terms of a displacement boundary condition using the polynomial function $\ol{u}(t) = \frac{2^{8}\ol{u}_{\text{max}}} { {T}^{8}} (t)^4(t-T)^4$, where $\ol{u}_{\text{max}}$ is the amplitude of the impact wave and $T$ the duration in which the load is applied.  Initially, the bar is at rest. 
The problem will be solved using both, a standard single-scale finite element problem referred to as direct numerical simulation (DNS), as well as the proposed dynamic FE$^2$ framework, c.f. Figure \ref{fig:DNS_FE2} respectively. 
The DNS discretizes the microscopic phases at the macroscale into a large number of finite elements with a length of $l_E$. 
It thereby serves as overkill reference for the multiscale approach. 
The FE$^2$ simulations have a macroscopic element length of ${l}_{\ol{\text{{E}}}}$ and make use of the same element length $l_E$ at the microscale for better {direct} comparability {of the microscopic fields to the DNS}. 
To approximate the displacement fields of the elements, linear shape functions and two Gauss points are used for all scales. 
As shown in Figure~\ref{fig:DNS_FE2}(b), each microscopic RVE calculation is associated to a single macroscopic integration point. 
The corresponding parameters for each simulation, regarding geometry, material parameters and loading will be listed in the caption of each figure. 
\begin{figure}[b!]
	\centering
	\begin{subfigure}{.55\textwidth}
		\centering
		\tikzset{external/export next=false}
\begin{tikzpicture}


\begin{axis}[
width=8.0cm,height=7.0cm,
,xlabel={\footnotesize Coordinate $\ol{{X}}_1$ in mm},
ylabel={\footnotesize Displacement $\ol{u}_1(\ol{X})$ in mm},
grid=none,
/pgf/number format/.cd,
set decimal separator={.},
1000 sep={},
precision=5,
legend style={at={(1,0.999)},anchor=north east},
legend style={draw=none},
every tick label/.append style={font=\footnotesize },
cycle list name=style-colors,
xmin=0,xmax=10000,
]
\addplot plot[graph_greenish,mark=|,only marks] table[x index=0,y index=1] {\tikzpath/convergence/x_plot_dat00300_reference_reduced-30.dat};
\addplot plot[graph_green, mark=none, thick] table[x index=0,y index=1] {\tikzpath/convergence/x_plot_dat00300_fe2_reduced-2.dat};
\addplot plot[graph_orange,mark=|, densely dashed,only marks] table[x index=0,y index=1] {\tikzpath/convergence/x_plot_dat00600_reference_reduced-30.dat};
\addplot plot[graph_red,mark=none, thick] table[x index=0,y index=1] {\tikzpath/convergence/x_plot_dat00600_fe2_reduced-2.dat};
\addplot plot[graph_gray,mark=|, only marks] table[x index=0,y index=1] {\tikzpath/convergence/x_plot_dat00900_reference_reduced-30.dat};
\addplot plot[graph_blue,mark=none, thick,solid] table[x index=0,y index=1] {\tikzpath/convergence/x_plot_dat00900_fe2_reduced-2.dat};

\legend{\tiny $t = 0.015$s DNS,\tiny $t = 0.015$s  FE$^2$,\tiny $t = 0.030$s DNS,\tiny $t = 0.030$s FE$^2$,\tiny $t = 0.045$s DNS,\tiny $t = 0.045$s FE$^2$}

\end{axis}

\end{tikzpicture}
		\caption{}
		\label{fig:convergenceA}
	\end{subfigure}%
	\hfill
	\begin{subfigure}{.45\textwidth}
		\centering
		\tikzset{external/export next=false}
\begin{tikzpicture}


\begin{axis}[
width=6.8cm,height=7cm,
,xlabel={\footnotesize Number of Iterations},
ylabel={\footnotesize $\left| \Delta \bD^* \right| $},
grid=none,
/pgf/number format/.cd,
set decimal separator={.},
1000 sep={},
precision=5,
ymode=log,
legend style={legend pos=south west},
every tick label/.append style={font=\footnotesize },
cycle list name=tu-colors,
xmin=1,xmax=4,
ymin=0,
xtick={1,2,3,4},
]
\addplot [dashed,domain=1:4,forget plot]{1e-8};
\addplot plot[graph_green,mark=square*, thick] table[x index=0,y index=2] {\tikzpath/convergence/xx_convergence_0300.dat};
\addplot plot[graph_red,mark=triangle*, thick] table[x index=0,y index=2] {\tikzpath/convergence/xx_convergence_0600.dat};
\addplot plot[graph_blue,mark=diamond*, thick] table[x index=0,y index=2] {\tikzpath/convergence/xx_convergence_0900.dat};
\pgfplotsset{cycle list shift=1}

\legend{\footnotesize $t = 0.015$s,\footnotesize $t = 0.030$s,\footnotesize $t = 0.045$s}

\end{axis}

\end{tikzpicture}
		\caption{}
		\label{fig:convergenceB}
	\end{subfigure}
	\caption{Analysis of algorithmic consistency: 
		(a) Comparison of displacement fields and (b) Convergence of the macroscopic Newton iteration with a tolerance of $10^{-8}$. 
		The simulation parameters are $L = 10000$~mm, $l_M = 10$~mm, $l_{\ol{\text{E}}} = 33.33$~mm, $\ol{u}_{\text{max}}~=~100$~mm, $T = 0.01$~s, $\Delta t = 5\cdot 10^5$~s and basic unit cell type A as RVE. }
	\label{fig:convergence}
\end{figure}
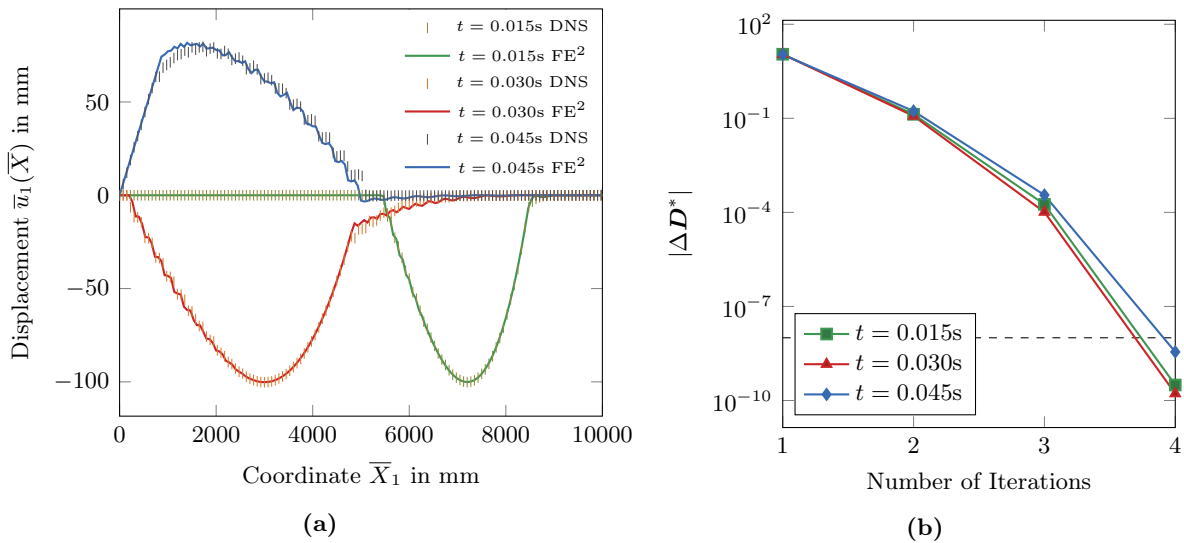
\subsection{Consistency of Numerical Framework}
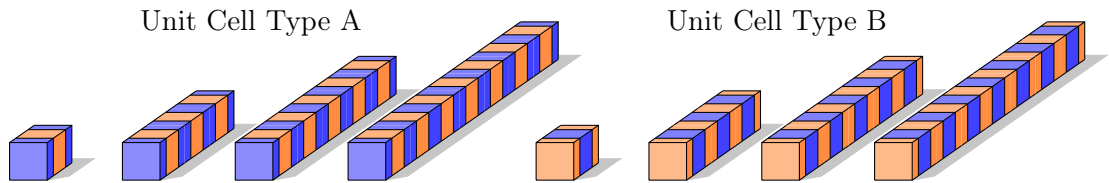
\begin{figure}[b!]
	\centering
	\setlength{\unitlength}{1 cm} 
\begin{picture}(14,2.5)

\put(  0,0.3){
\put(  2,2){Unit Cell Type A}
\put(  9,2){Unit Cell Type B}
\put( 0,0){
	\tikzset{external/export next=false}	
	\begin{tikzpicture}
	\drawRVE{1}{0.5}{0.4}{35}{0.6}{0.3}{orange}{blue}
	\end{tikzpicture}
}
\put( 1.5,0){
\tikzset{external/export next=false}	
\begin{tikzpicture}
\drawRVE{3}{0.5}{0.4}{35}{0.6}{0.3}{orange}{blue}
\end{tikzpicture}
}
\put( 3,0){
\tikzset{external/export next=false}	
\begin{tikzpicture}
\drawRVE{5}{0.5}{0.4}{35}{0.6}{0.3}{orange}{blue}
\end{tikzpicture}
}
\put( 4.5,0){
	\tikzset{external/export next=false}	
	\begin{tikzpicture}
	\drawRVE{7}{0.5}{0.4}{35}{0.6}{0.3}{orange}{blue}
	\end{tikzpicture}
}

\put( 7,0){
	\tikzset{external/export next=false}	
	\begin{tikzpicture}
	\drawRVE{1}{0.5}{0.4}{35}{0.6}{0.3}{blue}{orange}
	\end{tikzpicture}
}
\put( 8.5,0){
	\tikzset{external/export next=false}	
	\begin{tikzpicture}
	\drawRVE{3}{0.5}{0.4}{35}{0.6}{0.3}{blue}{orange}
	\end{tikzpicture}
}
\put( 10,0){
	\tikzset{external/export next=false}	
	\begin{tikzpicture}
	\drawRVE{5}{0.5}{0.4}{35}{0.6}{0.3}{blue}{orange}
	\end{tikzpicture}
}
\put( 11.5,0){
	\tikzset{external/export next=false}	
	\begin{tikzpicture}
	\drawRVE{7}{0.5}{0.4}{35}{0.6}{0.3}{blue}{orange}
	\end{tikzpicture}
}
}
\end{picture}
	\caption{Selection of RVE choices with different numbers of basic unit cells. 
		\label{fig:RVEchoices}}
\end{figure}

This example analyzes the convergence behavior of the macroscopic Newton iteration. 
In Figure~\ref{fig:convergence}(a), the distribution of macroscopic displacement fields is shown at three different time instances for both, the DNS and the FE$^2$ calculation. 
As RVE, the basic unit cell of the type A (c.f. Figure \ref{fig:RVEchoices}) is used. 
It can be seen, that the dynamic multiscale framework approximates the overall behavior well and even captures some of the smaller waves arising from the microstructure. 
A better representation of the wave propagation might be achieved by using finer time steps, but this would generally make the calculation converge faster as the initial values are already closer to the solution, defying the objective to properly test the tangent moduli. 
The convergence behavior for the three arbitrarily chosen time frames is depicted in Figure \ref{fig:convergence}(b). Quadratic convergence of the norm of the updates of the nodal displacements $\left| \Delta \bD^* \right|$ is observed. 
This demonstrates that the macroscopic tangent moduli, incorporating both the microscale inertia forces as well as possible constraints, have been derived in a consistent manner. 
%
%

\begin{figure}[b!]
	\centering
	\setlength{\unitlength}{1 cm} 
\begin{picture}(16.0,7.3)
\put(  0.0, 0.3){

\tikzset{external/export next=false}
\begin{tikzpicture}


\begin{axis}[
width=15.0cm,height=7.0cm,
,xlabel={\footnotesize Coordinate $\ol{{X}}_1$ in mm},
ylabel={\footnotesize Displacement $\ol{u}_1(\ol{X})$ in mm},
grid=none,
/pgf/number format/.cd,
set decimal separator={.},
1000 sep={},
precision=5,
legend style={at={(1,0.99)},anchor=north east},
legend style={draw=none},
every tick label/.append style={font=\footnotesize },
cycle list name=style-colors,
xmin=0,xmax=10000,
ymin=-19,ymax=100,
]
\addplot [solid,domain=0:10000,forget plot]{0};
\addplot plot[mark=none, thick,solid] table[x index=0,y index=1] {\tikzpath/compareAB/x_plot_dat00900_fe2_RVEA.dat};
\addplot plot[mark=none, thick,solid] table[x index=0,y index=1] {\tikzpath/compareAB/x_plot_dat00900_fe2_RVEB.dat};

\legend{\footnotesize FE$^2$ - Basic Unit Cell A, \footnotesize FE$^2$ - Basic Unit Cell B}

\end{axis}

\end{tikzpicture}
}
	
	\put( 6.05,4.8){
		\tikzset{external/export next=false}	
		\begin{tikzpicture}
		\drawRVE{1}{0.7}{0.9}{23}{0.0}{0.3}{blue}{orange}
		
		\coordinate (P1) at (-1,.1);
		\coordinate (P2) at (-0.1,0.35);
		\node[circle,draw=black, inner sep=0pt,minimum size=5pt] (N1) at (P1) {};
		\draw[->] (N1)--(P2);

		\end{tikzpicture}
	}

	\put( 7.2,3){
		\tikzset{external/export next=false}	
		\begin{tikzpicture}
		\drawRVE{1}{0.7}{0.9}{23}{0.0}{0.3}{orange}{blue}

		\coordinate (P1) at (-1.8,.25);
		\coordinate (P2) at (-0.1,0.35);
		\node[circle,draw=black, inner sep=0pt,minimum size=5pt] (N1) at (P1) {};
		\draw[->] (N1)--(P2);

		\end{tikzpicture}
	}
	
\end{picture}
	\caption{Comparison of macroscopic displacement fields for two different RVEs with different basic unit cells types.  The simulation parameters are $L = 10000$ mm, $l_M~=~10$~mm, $l_{\ol{\text{E}}}~=~33.33$~mm, $\ol{u}_{\text{max}} = 100$~mm, $T = 0.01$ s, $t=0.045$ s and $\Delta t = 5\cdot 10^5$ s. }
	\label{fig:compareDisplacementAB}
\end{figure}
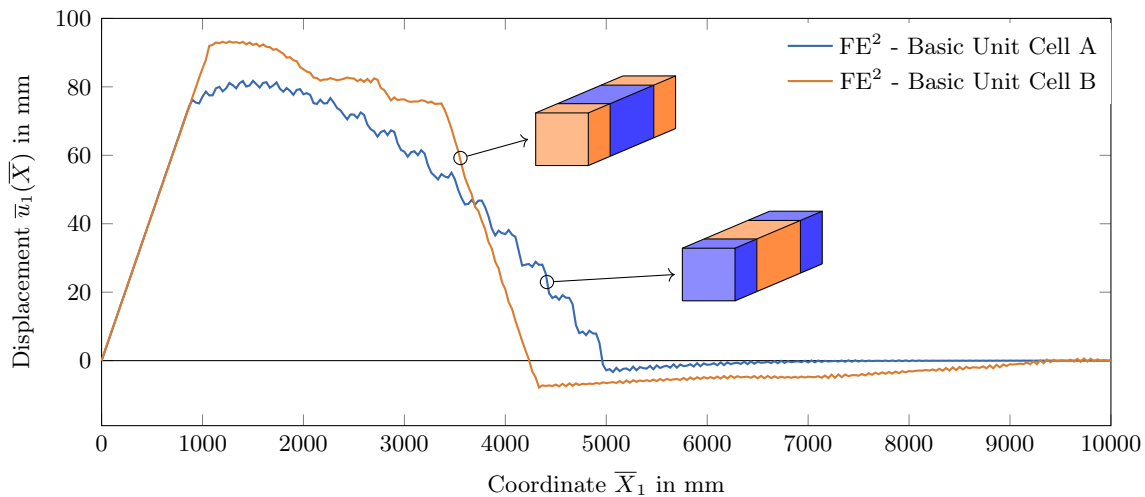
\subsection{Analysis of the Unit Cell Concept under Dynamic Loading}
\begin{figure}[t]
	\centering
	\begin{subfigure}[b]{.45\textwidth}
		\centering
		\setlength{\unitlength}{1 cm} 
\begin{picture}(8.0,6.3)
\put(  0.0, 0.3){
	
\tikzset{external/export next=false}
\begin{tikzpicture}

\begin{axis}[
width=7.4cm,height=6cm,
,xlabel={\footnotesize Basic Unit Cells per RVE},
ylabel={\footnotesize Error $\epsilon_{\text{time}}$ in mm},
grid=none,
/pgf/number format/.cd,
set decimal separator={.},
1000 sep={},
precision=5,
legend style={at={(1,0.99)},anchor=north east},
legend style={draw=none},
cycle list name=style-colors,
every tick label/.append style={font=\footnotesize },
xmin=1,xmax=7,
xtick={1,3,5,7},
]
\addplot plot[mark=*,thick] table[x index=0,y index=1] {\tikzpath/errormeasure/compareAB5.dat};


\end{axis}

\end{tikzpicture}
}

\put( 1.2,4.7){
	\put( 0,0){
		\tikzset{external/export next=false}	
		\begin{tikzpicture}
		\drawRVE{1}{0.25}{0.4}{45}{0.0}{0.3}{orange}{blue}
		\end{tikzpicture}
	}
	\put(0.3,-0.1){
		\tikzset{external/export next=false}	
		\begin{tikzpicture}
		\drawRVE{1}{0.25}{0.4}{45}{0.0}{0.3}{blue}{orange}
		\end{tikzpicture}
	}
}

\put( 3.3,4.3){
	\put( 0,0){
		\tikzset{external/export next=false}	
		\begin{tikzpicture}
		\drawRVE{3}{0.25}{0.4}{45}{0.0}{0.3}{orange}{blue}
		\end{tikzpicture}
	}
	\put(0.3,-0.1){
		\tikzset{external/export next=false}	
		\begin{tikzpicture}
		\drawRVE{3}{0.25}{0.4}{45}{0.0}{0.3}{blue}{orange}
		\end{tikzpicture}
	}
}

\put( 4.5,3.2){
	\put( 0,0){
		\tikzset{external/export next=false}	
		\begin{tikzpicture}
		\drawRVE{5}{0.25}{0.4}{45}{0.0}{0.3}{orange}{blue}
		\end{tikzpicture}
	}
	\put(0.3,-0.1){
		\tikzset{external/export next=false}	
		\begin{tikzpicture}
		\drawRVE{5}{0.25}{0.4}{45}{0.0}{0.3}{blue}{orange}
		\end{tikzpicture}
	}
}

\put( 6,1.6){
	\put( 0,0){
	\tikzset{external/export next=false}	
	\begin{tikzpicture}
	\drawRVE{7}{0.25}{0.4}{45}{0.0}{0.3}{orange}{blue}
	\end{tikzpicture}
	}
\put(0.3,-0.1){
\tikzset{external/export next=false}	
\begin{tikzpicture}
\drawRVE{7}{0.25}{0.4}{45}{0.0}{0.3}{blue}{orange}
\end{tikzpicture}
}

}

\end{picture}
		\caption{}
		\label{fig:compareErrorAB}
	\end{subfigure}%
	\hfill
	\begin{subfigure}[b]{.45\textwidth}
		\centering
		\tikzset{external/export next=false}
\begin{tikzpicture}


\begin{axis}[
width=7.4cm,height=6cm,
,xlabel={\footnotesize Basic Unit Cells per RVE},
ylabel={\footnotesize Error $\epsilon_{\text{time}}$ in mm},
grid=none,
/pgf/number format/.cd,
set decimal separator={.},
1000 sep={},
precision=5,
legend style={at={(1,0.99)},anchor=north east},
legend style={draw=none},
cycle list name=style-colors,
every tick label/.append style={font=\footnotesize },
xmin=1,xmax=7,
ymin=0.035,
ymax=0.105,
xtick={1,3,5,7},
]
\addplot plot[mark=square*,thick] table[x index=0,y index=1] {\tikzpath/errormeasure/plot_RVEA_lgr.dat};
\addplot plot[mark=triangle*, thick] table[x index=0,y index=1] {\tikzpath/errormeasure/plot_RVEB_lgr.dat};

\legend{\footnotesize Unit Cells Type A, \footnotesize Unit Cells Type B}

\end{axis}

\end{tikzpicture}
		\caption{}
		\label{fig:compareErrorABtoDNS}
	\end{subfigure}
	\caption{
		Analysis of different RVE choices: 
		(a) Direct comparison of RVEs with unit cell type A and B shown by with increasing number of basic unit cells per RVE (the error is computed as difference between the response of the two unit cell types, not with respect to the DNS). 
		(b) Error of unit cell type A and B, compared to DNS as reference. 
		The simulation parameters are $L = 10000$ mm, $l_M = 2.5$ mm, $l_{\ol{\text{E}}} = 20$ mm, $\ol{u}_{\text{max}} = 100$ mm, $T = 0.01$ s, $\Delta t = 5\cdot 10^5$ s, $n_{\text{timesteps}} = 400$. 
		\label{fig:compareError}}
\end{figure}
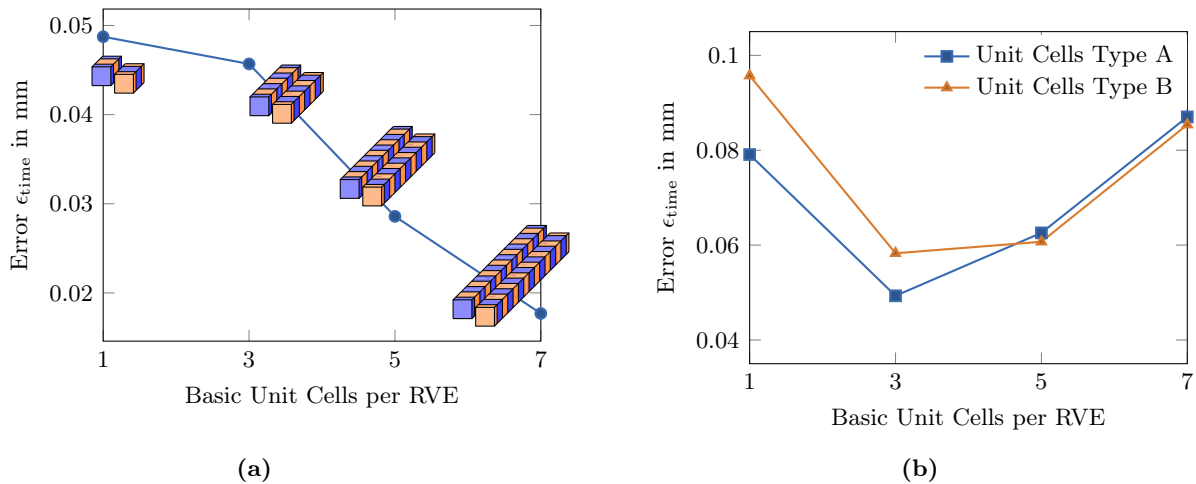

For quasi-static homogenization simulations of periodic microstructures, it is known that the resulting macroscopic answer, as well as the corresponding microscopic fields are invariant with respect to the specific choice of unit cell, as long as an admissible periodic unit cell is chosen. 
In contrast to quasi-static cases, the distribution of the mass relative to the geometrical center matters in a dynamic setting. 
An extreme example is shown in Figure~\ref{fig:compareDisplacementAB}, which compares the macroscopic displacement field at $t=0.045$ s presented in the first example in Figure \ref{fig:convergenceA} with a simulation using the basic unit cell type B as an RVE (c.f. Figure \ref{fig:RVEchoices}). 
To properly measure the influence of different RVE choices on the FE$^2$ simulation, an objective error measure $\epsilon$ is considered. 
It is defined as average difference of the macroscopic displacement fields $ \epsilon = \sum_i^{n_{\text{nodes}}}\left| \ol{u}^{\text{I}}_{i}(t_j) - \ol{u}^{\text{II}}_{i}(t_j) \right|/n_{\text{nodes}}$. 
This measure can be evaluated for each time step and thus, the average is once more computed over the number of time steps $\epsilon_{\text{time}}~=~\sum_j^{n_{\text{timesteps}}} \epsilon_j/n_{\text{timesteps}}$. 
Figure~\ref{fig:compareError} shows the calculated errors of different choices of the RVE. 
For the comparison, RVEs with multiple periods of the same unit cell type, as depicted in Figure~\ref{fig:RVEchoices}, were considered. 
Two effects can be observed:
The first, presented in Figure~\ref{fig:compareErrorAB}, is that the difference in macroscopic displacements between different choices of unit cell type decreases, when the number of unit cells per RVE is increased. 
This means that the choice of particular basic unit cell type does not matter as long as the RVE is chosen large enough. 
The second effect, shown in Figure~\ref{fig:compareErrorABtoDNS}, is that the error, computed as difference to the DNS reference, increases when the size of the RVE, relative to the macroscopic element length, gets too large. 
Then, errors resulting from a violation of the scale separation assumption are obtained. 
Generally, the second effect can be neglected, as calculations with RVE sizes larger than the macroscopic element length have little practical application when using FE${}^2$. 
At this point it is favorable to use domain decomposition approaches instead of a homogenization method in order to avoid the scale separation assumption. 
%
\begin{table}[b]
	\begin{center}
		\begin{tabular}{cc|cccc} \toprule
			\multicolumn{2}{c|}{Number of Unit Cells per RVE}&1&3&5&7\\
			\midrule		
			Unit Cell Type& $\ol{\bu}$-Link& \multicolumn{4}{c}{Number of Time Steps}\\
			\midrule		
			\multirow{2}{*}{A}&$\ol{\bu} = \left\langle \bu \right\rangle$&$1000$&$940$&$1000$&$1000$\\
			&$\widetilde{\bu}_{\text{corner}} = \bzero$&$1000$&$671$&$634$&$1000$\\
			\midrule		
			\multirow{2}{*}{B}&$\ol{\bu} = \left\langle \bu \right\rangle$&$1000$&$456$&$1000$&$1000$\\
			&$\widetilde{\bu}_{\text{corner}} = \bzero$&$944$&$420$&$192$&$166$\\
			\bottomrule	
		\end{tabular}
	\end{center}
	\caption{Number of time steps before either the simulation crashed (divergence of Newton iteration at microscale) or the intended complete set of 1000 time steps was successfully reached. 
		Different choices of RVEs and constraints were analyzed.\label{tab:ncalc}}
\end{table}
\subsection{Influence of Displacement Constraints}
Finally the proposed displacement link $\ol{\bu} = \left\langle \bu \right\rangle$ is analyzed for the examples in Figure \ref{fig:compareErrorABtoDNS}, in comparison to the standard displacement link for quasi-static periodic homogenization, where the fluctuations at the RVE corner nodes is set to zero, i.e. $\widetilde{\bu}_{\text{corner}} = \bzero$. 
For the quasi-1D example analyzed here, this is equivalent to setting the integral over the surface equal to the corresponding macroscopic displacements, which has been taken into account in other dynamic homogenization schemes.\\ 
The first observation is, that using the proposed volume constraint $\ol{\bu} = \left\langle \bu \right\rangle$ results in a more robust framework in terms of stability of the Newton-Raphson iterations. 
Furthermore, slightly smaller error values are obtained compared to the DNS reference. 
Table~\ref{tab:ncalc} shows the number of time steps reached before either the calculations crashed (due to diverging Newton iterations at the microscale) or they were finished successfully after the intended complete set of 1000 time steps. 
Especially the calculations using the unit cell type B in combination with the zero fluctuations of the corner nodes, underperformed the other scenarios. 
To understand the difference between the performance of the displacement links, it is necessary to examine the behavior at the RVE level. \\
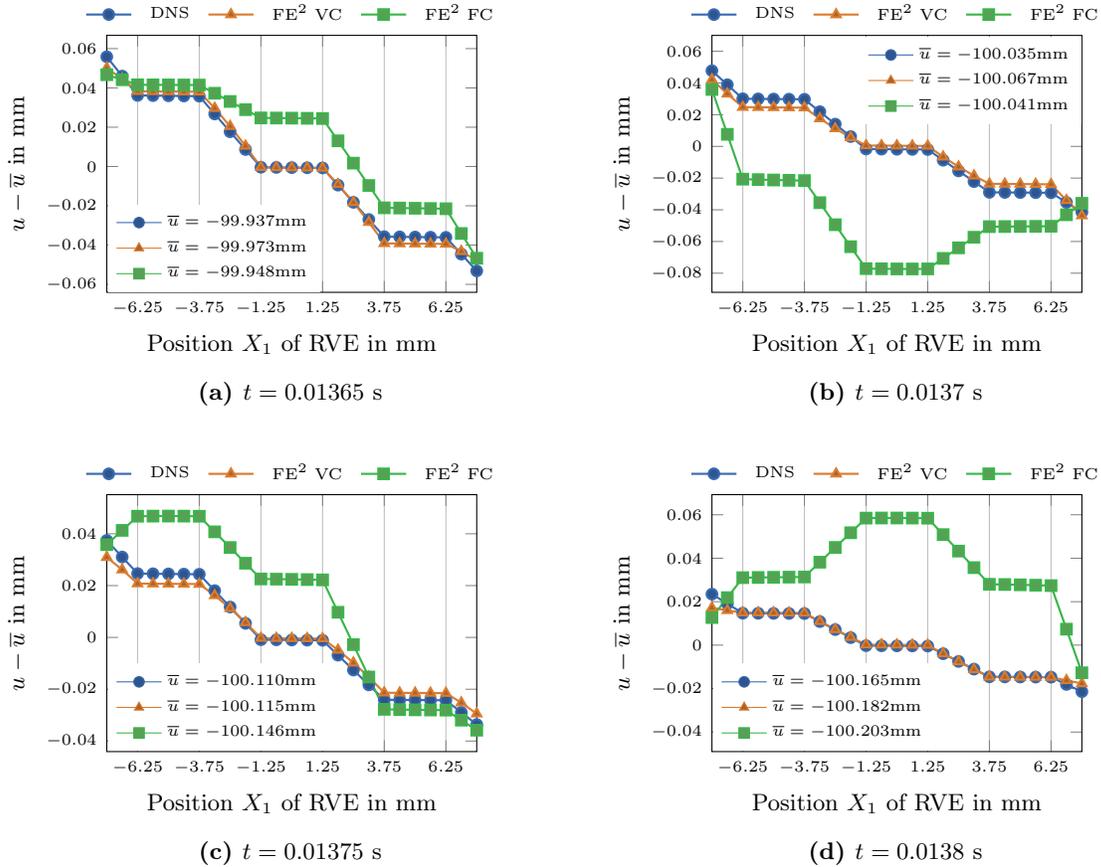
\begin{figure}[b!]
	\centering
	\begin{subfigure}{.5\textwidth}	
		\centering
		\setlength{\unitlength}{1 cm} 
		\begin{picture}(8,5)
		\put(0,0){\tikzset{external/export next=false}
\begin{tikzpicture}


\begin{axis}[
width=6.5cm,height=5cm,
xlabel={\footnotesize Position $X_1$ of RVE in mm},
ylabel={\footnotesize $u - \ol{u}$ in mm},
grid=none,
/pgf/number format/.cd,
yticklabel style={/pgf/number format/fixed},
set decimal separator={.},
1000 sep={},
legend style={at={(0.5,1.0)},anchor=south},
legend style={draw=none,column sep=1ex},
legend columns=-1,
cycle list name=style-colors,
every tick label/.append style={font=\tiny },
xmin=-7.5,xmax=7.5,
xmajorgrids,
xtick={-6.25,-3.75,-1.25,6.25,3.75,1.25},
ytick={-0.06,-0.04,-0.02,0.0,0.02,0.04,0.06},
]
\addplot plot[mark=*,thick] table[x index=0,y index=2] {\tikzpath/microscale/DNS_plot_dat00273.dat};
\addplot plot[mark=triangle*,thick] table[x index=0,y index=2] {\tikzpath/microscale/lgr_plot_dat00273.dat};
\addplot plot[mark=square*,thick] table[x index=0,y index=3] {\tikzpath/microscale/fix_plot_dat00273.dat};

\legend{\tiny  DNS,\tiny FE$^2$  VC,\tiny FE$^2$ FC };
\end{axis}

\end{tikzpicture}}
		\put(1.45,0.95){\tikzset{external/export next=false}
\begin{tikzpicture}
\begin{customlegend}[legend style={draw=none},legend entries={\tiny $\ol{u} =-99.937$mm,\tiny $\ol{u} = -99.973$mm,\tiny $\ol{u} = -99.948$mm}]
\addlegendimage{graph_blue,fill=.!80!black,mark=*}
\addlegendimage{graph_orange,fill=.!80!black,mark=triangle*}
\addlegendimage{rub_green,fill=.!80!black,mark=square*}
\end{customlegend}
\end{tikzpicture}}
		\end{picture}
		\caption{$ t = 0.01365$ s}
		\label{fig:microscalet1}
	\end{subfigure}%
	\begin{subfigure}{.5\textwidth}
		\centering
		\setlength{\unitlength}{1 cm} 
		\begin{picture}(8,5)
		\put(0,0){\tikzset{external/export next=false}
\begin{tikzpicture}


\begin{axis}[
width=6.5cm,height=5cm,
xlabel={\footnotesize Position $X_1$ of RVE in mm},
ylabel={\footnotesize $u - \ol{u}$ in mm},
grid=none,
/pgf/number format/.cd,
yticklabel style={/pgf/number format/fixed},
set decimal separator={.},
1000 sep={},
legend style={at={(0.5,1.0)},anchor=south},
legend style={draw=none,column sep=1ex},
legend columns=-1,
cycle list name=style-colors,
every tick label/.append style={font=\tiny },
xmin=-7.5,xmax=7.5,
ymax=0.07,
xmajorgrids,
xtick={-6.25,-3.75,-1.25,6.25,3.75,1.25},
ytick={-0.1,-0.08,-0.06,-0.04,-0.02,0.0,0.02,0.04,0.06},
]
\addplot plot[mark=*,thick] table[x index=0,y index=2] {\tikzpath/microscale/DNS_plot_dat00274.dat};
\addplot plot[mark=triangle*,thick] table[x index=0,y index=2] {\tikzpath/microscale/lgr_plot_dat00274.dat};
\addplot plot[mark=square*,thick] table[x index=0,y index=3] {\tikzpath/microscale/fix_plot_dat00274.dat};

\legend{\tiny  DNS,\tiny FE$^2$  VC,\tiny FE$^2$ FC, }

\end{axis}

\end{tikzpicture}}
		\put(3.4,3.18){\tikzset{external/export next=false}
\begin{tikzpicture}

\begin{customlegend}[legend style={draw=none},legend entries={\tiny $\ol{u} =-100.035$mm,\tiny $\ol{u} = -100.067$mm,\tiny $\ol{u} = -100.041$mm}]

\addlegendimage{graph_blue,fill=.!80!black,mark=*}

\addlegendimage{graph_orange,fill=.!80!black,mark=triangle*}

\addlegendimage{rub_green,fill=.!80!black,mark=square*}

\end{customlegend}

\end{tikzpicture}}
		\end{picture}
		\caption{$ t = 0.0137$ s}
		\label{fig:microscalet2}
	\end{subfigure}
	\vskip\baselineskip
	\centering
	\begin{subfigure}{.5\textwidth}	
		\centering
		\setlength{\unitlength}{1 cm} 
		\begin{picture}(8,5)
		\put(0,0){\tikzset{external/export next=false}
\begin{tikzpicture}


\begin{axis}[
width=6.5cm,height=5cm,
xlabel={\footnotesize Position $X_1$ of RVE in mm},
ylabel={\footnotesize $u - \ol{u}$ in mm},
grid=none,
/pgf/number format/.cd,
yticklabel style={/pgf/number format/fixed},
set decimal separator={.},
1000 sep={},
legend style={at={(0.5,1.0)},anchor=south},
legend style={draw=none,column sep=1ex},
legend columns=-1,
cycle list name=style-colors,
every tick label/.append style={font=\tiny },
xmin=-7.5,xmax=7.5,
ytick={-0.06,-0.04,-0.02,0.0,0.02,0.04,0.06},
xmajorgrids,
xtick={-6.25,-3.75,-1.25,6.25,3.75,1.25},
]
\addplot plot[mark=*,thick] table[x index=0,y index=2] {\tikzpath/microscale/DNS_plot_dat00275.dat};
\addplot plot[mark=triangle*,thick] table[x index=0,y index=2] {\tikzpath/microscale/lgr_plot_dat00275.dat};
\addplot plot[mark=square*,thick] table[x index=0,y index=3] {\tikzpath/microscale/fix_plot_dat00275.dat};

\legend{\tiny  DNS,\tiny FE$^2$  VC,\tiny FE$^2$ FC, }

\end{axis}

\end{tikzpicture}}
		\put(1.45,0.95){\tikzset{external/export next=false}
\begin{tikzpicture}

\begin{customlegend}[legend style={draw=none},legend entries={\tiny $\ol{u} =-100.110$mm,\tiny $\ol{u} = -100.115$mm,\tiny $\ol{u} = -100.146$mm}]

\addlegendimage{graph_blue,fill=.!80!black,mark=*}

\addlegendimage{graph_orange,fill=.!80!black,mark=triangle*}

\addlegendimage{rub_green,fill=.!80!black,mark=square*}

\end{customlegend}

\end{tikzpicture}}
		\end{picture}
		\caption{$ t = 0.01375$ s}
		\label{fig:microscalet3}
	\end{subfigure}%
	\begin{subfigure}{.5\textwidth}
		\centering
		\setlength{\unitlength}{1 cm} 
		\begin{picture}(8,5)
		\put(0,0){\tikzset{external/export next=false}
\begin{tikzpicture}


\begin{axis}[
width=6.5cm,height=5cm,
xlabel={\footnotesize Position $X_1$ of RVE in mm},
ylabel={\footnotesize $u - \ol{u}$ in mm},
grid=none,
/pgf/number format/.cd,
yticklabel style={/pgf/number format/fixed},
set decimal separator={.},
1000 sep={},
legend style={at={(0.5,1.0)},anchor=south},
legend style={draw=none,column sep=1ex},
legend columns=-1,
cycle list name=style-colors,
every tick label/.append style={font=\tiny },
xmin=-7.5,xmax=7.5,
ymin=-0.049,
xmajorgrids,
xtick={-6.25,-3.75,-1.25,6.25,3.75,1.25},
ytick={-0.06,-0.04,-0.02,0.0,0.02,0.04,0.06},
]
\addplot plot[mark=*,thick] table[x index=0,y index=2] {\tikzpath/microscale/DNS_plot_dat00276.dat};
\addplot plot[mark=triangle*,thick] table[x index=0,y index=2] {\tikzpath/microscale/lgr_plot_dat00276.dat};
\addplot plot[mark=square*,thick] table[x index=0,y index=3] {\tikzpath/microscale/fix_plot_dat00276.dat};

\legend{\tiny  DNS,\tiny FE$^2$  VC,\tiny FE$^2$ FC, }

\end{axis}

\end{tikzpicture}}
		\put(1.45,0.95){\tikzset{external/export next=false}
\begin{tikzpicture}

\begin{customlegend}[legend style={draw=none},legend entries={\tiny $\ol{u} =-100.165$mm,\tiny $\ol{u} = -100.182$mm,\tiny $\ol{u} = -100.203$mm}]

\addlegendimage{graph_blue,fill=.!80!black,mark=*}

\addlegendimage{graph_orange,fill=.!80!black,mark=triangle*}

\addlegendimage{rub_green,fill=.!80!black,mark=square*}

\end{customlegend}

\end{tikzpicture}}
		\end{picture}
		\caption{$ t = 0.0138$ s}
		\label{fig:microscalet4}
	\end{subfigure}
	\caption{Comparison of microscopic displacement fields obtained from the FE$^2$ simulations with the DNS. 
		The displacements have been normalized by {$\ol{\bu}$ (for the DNS, average displacement)} to analyze the quality of microscopic displacements more or less independent {of} the macroscopic displacements. 
		The two different displacement links, the volume constraint (VC) and the fixed corners (FC) are analyzed. 
		The simulation parameters are $L = 10000$~mm, $l_M = 2.5$~mm, $l_{\ol{\text{E}}} = 20$~mm, $\ol{u}_{\text{max}} = 100$~mm, $T = 0.01$~s, $\Delta t = 5\cdot 10^5$~s, location of the macroscale integration point $\ol{X}_1 = 7504,23$~mm, section of the DNS displacement field from $\ol{X}_1 = 7496.25$~mm to $\ol{X}_1 = 7511.25$~mm, RVE with three periods of basic unit cell types B. }
	\label{fig:microscale}
\end{figure}
{Here the examples with the RVEs consisting of three periods of the basic unit cell B are further analyzed.}
Figure~\ref{fig:microscale} compares the microscopic displacements for four relevant time instances right before the peak of the input wave passes through the RVEs. 
More specifically, the differences between the microscopic displacement fields $\bu$ of an RVE and the respective macroscopic displacements $\ol{\bu}$. 
To compare the DNS, an effective $\ol{\bu}$ has been computed as the average displacement over the associated length.
Thereby, the quality of the microscopic fields can be analyzed independently from the macroscopic displacements. 
{With this} , the two different displacement constraint options can be effectively compared with the reference solution obtained from DNS. 
The graphs show, that the fixed corner constraint leads to artificially increased displacement intensities at the microscale due to the constricted boundary. 
These increased displacements eventually lead to extreme deformations in single elements at the microscale, crashing the simulation. 
The proposed displacement volume constraint however, leads to a softer constraint which results in a more robust computation while still enabling dynamic effects which agree well with the ones from the reference DNS. 
In the presented examples, the only rate-dependent influence are inertia forces. 
In cases where also rate-dependent material properties are included we expect the influence of different displacement constraints on the overall simulation to increase, in favor for the proposed volume constraint.


\FloatBarrier
\section{Conclusion} \label{sec:conclusion}
In this paper, a general purpose, consistent, two-scale homogenization framework for dynamics at the macro- and microscale was proposed in the sense of the FE${}^2$ method. 
The framework does not include any simplifications such as linearized strains, explicit time integration or partly quasi-static scenarios. 
The only assumption taken into account is a sufficiently pronounced scale separation, which is anyway essential requirement of any FE${}^2$ approach. 
Therefore, it enables the simulation of various structural problems of complex micro-heterogeneous materials under dynamic loading such as impact. 
The main aspects are: (i) the incorporation of the complete balance of momentum at the micro- and macroscale including inertia forces, (ii) a large-strain formulation enabling the simulation of a wide range of macro- and micro-mechanical phenomena, (iii) extended microscopic boundary conditions resulting in a more robust scheme and giving the possibility for non-periodic RVE boundaries, and (iv) the derivation of consistent macroscopic tangent moduli ensuring quadratically converging macroscopic iterations. 
The presented results on different choices of RVEs show that  different admissible unit cells of a periodic microstructure lead also to a different mechanical response - even if periodic boundary conditions are employed. 
This is in contrast to quasi-static scenarios. 
However, these results should not entail that the most basic unit cell is necessarily a bad choice for a simulation, but the specific choice of this basic unit cell is not unique. 
Future work will focus on the analysis of more complex RVEs for non-periodic microstructures under dynamic loading. \par

\FloatBarrier

\section*{Acknowledgment}
The authors gratefully acknowledge financial support from the  German Research Foundation (DFG) within project B1 of the Training Research Group (GRK) 2250 ``Mineral-Bonded Composites for Enhanced Structural Impact Safety''. 
Furthermore, fruitful discussions with Celia Reina (University of Pennsylvania) are greatly appreciated. 

\newpage
\section*{Appendix}
\begin{appendix} \section{Matrix Abbreviations} \label{sec:appendixA}
\begin{table}[htb]
	\centering	
	\begin{tabular}{ @{}cl@{\hskip 1cm}l@{}  }
		\toprule
		\multicolumn{2}{c}{Global\hspace{1cm} }&\multicolumn{1}{c}{Element}\\
		\multicolumn{1}{c}{\small Definition}&\multicolumn{1}{c}{\small Size\hspace{1cm} }&\\
		\midrule
		$\displaystyle {{\bK}} = \A^{n_{\text{ele}}}_{e=1} \bk^{e}$&\small$n_{\text{edf}}\times n_{\text{edf}}$&$\displaystyle k^{e}_{PQ} = \int_{\B^{e}} B^{e}_{ijP}\IA_{ijkl}B^{e}_{klQ}\dV$\\[9pt]
		$\displaystyle \bL = \A^{n_{\text{ele}}}_{e=1} \bl^{e}$&\small$n_{\text{edf}}\times n_{\text{dm}}^{2}$&$\displaystyle l^{e}_{Pij} = \int_{\B^{e}} B^{e}_{klP}\IA_{klij}\dV $ \\[9pt]
		$\displaystyle \bM = \A^{n_{\text{ele}}}_{e=1}\bbm^{e}$&\small$n_{\text{edf}}\times n_{\text{edf}}$&$\displaystyle m^{e}_{PQ} = \int_{\B^{e}} N^{e}_{Pi} \rho_{0}N^{e}_{Qi}\dV$\\[9pt]
		$\displaystyle \bW = \A^{n_{\text{ele}}}_{e=1} \bw^{e}$&\small$n_{\text{edf}}\times n_{\text{dm}}$&$\displaystyle w^{e}_{Pi} = \int_{\B^{e}}\rho_{0} N^{e}_{Pi}\dV$\\[9pt]
		$\displaystyle \bZ = \A^{n_{\text{ele}}}_{e=1} \bz^{e}$&\small$n_{\text{edf}}\times n_{\text{dm}}^{2}$&$\displaystyle z^{e}_{Pij} = \int_{\B^{e}}\rho_{0} N^{e}_{Pi} X_{j}\dV$\\[9pt]
		$\displaystyle \bG = \A^{n_{\text{ele}}}_{e=1} \bg^{e}$&\small$n_{\text{edf}}\times n_{\text{dm}}$&$\displaystyle g^{e}_{Pi} = \int_{\B^{e}}N^{e}_{Pi}\dV$\\[9pt]
		$\displaystyle \bV $&\small$ n_{\text{dm}}^{2}\times  n_{\text{dm}}$&$\displaystyle V_{ijk} = \rho_{0}\delta_{ik}X_{j} $\\[9pt]
		$\displaystyle \IY $&\small $n_{\text{dm}}^{2}\times n_{\text{dm}}^{2}$& $\displaystyle \IY_{ijkl} = \rho_{0} \delta_{ik}X_{l}X_{j}$\\
		\bottomrule
	\end{tabular}
	\caption{Overview of the used fields at element and global level, with $n_{\text{edf}}:$ number of DOF at element level and $n_{\text{dm}}:$ spacial dimension.}
	\label{tab:m_overview}
	
\end{table}
\vspace{1cm}
\begin{table}[htb]
	\centering	
	\begin{tabular}{ @{}l@{\hskip 1cm}l@{}  }
		\toprule
		\multicolumn{1}{c}{Matrix\hspace{1cm} }&\multicolumn{1}{c}{Size}\\
		\midrule
		$\bD^{*}=\left[\begin{array}{c|c}	 \widetilde{\bD}^{\text{T}} &	\Blambda^{\text{T}} \end{array}\right]^{\text{T}}$&		
		\small $(n_{\text{edf}}+ n_{\text{lgr}})\times 1$\\[8pt]
		$\bL^{*}=\left[\begin{array}{c|c}	 \bL^{\text{T}}  + \frac{{\alpha}}{\Delta t^{2}} \bZ^{\text{T}}   & 	\bzero  \end{array}\right]^{\text{T}}$&
		\small $(n_{\text{edf}}+ n_{\text{lgr}})\times n_{\text{dm}}^2$\\[8pt]
		$\bL^{\ol{*}}=\left[\begin{array}{c|c}	 \bL^{\text{T}}  + \frac{\ol{\alpha}}{\Delta t^{2}} \bZ^{\text{T}}   & 	\bzero  \end{array}\right]^{\text{T}}$&		
		\small $(n_{\text{edf}}+ n_{\text{lgr}})\times n_{\text{dm}}^2$\\[8pt]
		$\bW^{*}=\left[\begin{array}{c|c}	 \bW^{\text{T}}   & 	\bzero  \end{array}\right]^\text{T}$&
		\small $(n_{\text{edf}}+ n_{\text{lgr}})\times n_{\text{lgr}}$\\[8pt]
		$\bK^{*}=\left[\begin{array}{c|c}	\bK + \frac{{\alpha}}{\Delta t^{2}} \bM   & 	\bG \\ \hline
		\bG^{\text{T}} & \bzero \end{array}\right]$&
		\small $(n_{\text{edf}}+ n_{\text{lgr}})\times (n_{\text{edf}}+ n_{\text{lgr}})$\\
		\bottomrule
	\end{tabular}
	\caption{Overview of the extended fields, with $n_{\text{edf}}:$ number of DOF at element level, $n_{\text{lgr}}:$ number of DOF of Lagrange constraint and $n_{\text{dm}}:$ spacial dimension.}
	\label{tab:m*_overview}
\end{table}


 \end{appendix}

\FloatBarrier
\newpage
\bibliographystyle{references_style}
\bibliography{references} 

\end{document}